%% file: main-arxiv.tex
\documentclass[11pt,a4paper]{article}
\input{preamble-arxiv}

\author{Mingyu Song \quad
  Yizhou Wang \quad
  Yuqing Kong\thanks{Corresponding author.}\\[0.5em]
  CFCS, School of Computer Science, Peking University\\
  \texttt{\{mingyu.song,yizhou.wang,yuqing.kong\}@pku.edu.cn}}
\date{}

\begin{document}
\maketitle
\input{abstract-arxiv}

\bibliographystyle{plainnat}
\input{body}

\end{document}

%% file: preamble-arxiv.tex
\usepackage[margin=1in]{geometry}
\usepackage{graphicx,xcolor,amssymb}
\input{preamble-common}
\usepackage{amsthm}
\usepackage[numbers,square,sort&compress]{natbib}
\usepackage[hidelinks]{hyperref}

\theoremstyle{plain}
\newtheorem{theorem}{Theorem}[section]
\newtheorem{proposition}[theorem]{Proposition}
\newtheorem{lemma}[theorem]{Lemma}
\newtheorem{corollary}[theorem]{Corollary}
\theoremstyle{definition}
\newtheorem{definition}[theorem]{Definition}
\newtheorem{example}[theorem]{Example}

\providecommand{\Description}[2][]{\ignorespaces}

%% file: preamble-common.tex
\usepackage{mathtools}
\usepackage{placeins}
\usepackage{pgfplots}
\usepgfplotslibrary{fillbetween}
\pgfplotsset{compat=1.18}
\definecolor{intervalbandblue}{HTML}{3C5488}
\definecolor{midpointorange}{HTML}{F39B7F}
\definecolor{meanred}{HTML}{E64B35}
\definecolor{mediangreen}{HTML}{00A087}
\definecolor{rawmedianbrown}{HTML}{7E6148}
\definecolor{phasepositive}{HTML}{4DBBD5}
\definecolor{phasenegative}{HTML}{F39B7F}
\definecolor{optimalred}{HTML}{E64B35}
\pgfplotsset{
 comparison axis/.style={
  width=\linewidth, height=5.8cm,
  xmin=1.7, xmax=10.3, xtick={2,4,6,8,10},
  xlabel={Number of experts $n$},
  tick label style={font=\small}, label style={font=\small},
  axis line style={black!55}, tick style={black!55},
  grid=major, grid style={black!9},
  legend style={font=\footnotesize, draw=none, fill=none,
    cells={anchor=west}, column sep=0.7em},
  clip=true,
 },
}

\newcommand{\E}{\mathbb E}
\newcommand{\PP}{\mathbb P}
\newcommand{\Var}{\operatorname{Var}}
\newcommand{\WS}{\mathrm{WS}}
\numberwithin{equation}{section}

\title{Robust Aggregation of Quantile Statistics: from Median to Interval}

%% file: abstract-arxiv.tex
\begin{abstract}

Quantile-based summaries, such as prediction intervals and medians, are commonly used in economic forecasting, climate science, and public health. For highly skewed samples, experimental evidence suggests that estimates of the mean can be biased toward the median, whereas estimates of the median are relatively accurate. This motivates aggregating explicitly elicited quantiles rather than reported means.

We study robust aggregation of quantile reports for bounded outcomes under squared loss, minimizing worst-case regret relative to an omniscient Bayesian aggregator who knows the information structure and observes all private signals. Without restrictions on the information structure, no rule improves on the constant forecast at the range midpoint in worst-case regret. We therefore consider weak substitutes, a diminishing-returns condition on the value of information, and characterize optimal aggregation rules under this condition.

For central prediction intervals with common nominal coverage $\alpha$, the optimal rule, which we call \emph{regularized simple average} for prediction intervals, shrinks the average reported interval midpoint toward the range midpoint, with a weight determined by $\alpha$ and the number $n$ of experts. This optimal rule strictly improves on the constant baseline in worst-case regret when $\alpha<1/\sqrt{n}$; at or above this threshold, the constant rule is optimal. The optimal worst-case regret is minimized at a unique interior coverage level, revealing a nonmonotonic relationship between coverage and aggregation performance. For median reports, \emph{regularized simple average} for medians similarly shrinks the average reported median toward the range midpoint, assigning weight $1/n$ to the reported average and strictly improving on the constant baseline in worst-case regret for every finite $n\ge2$.
\end{abstract}

%% file: body.tex
\input{sections/introduction}
\input{sections/related_work}

\input{sections/model}
\input{sections/aggregation_of_intervals}
\input{sections/aggregation_of_medians}
\input{sections/discussion}

\FloatBarrier
\bibliography{references}

\appendix
\input{appendices/weak_substitutes}
\input{appendices/interval_aggregation_proofs}
\input{appendices/simple_average}

%% file: sections/introduction.tex
\section{Introduction}\label{sec:introduction}

\input{figures/duke_questionnaire}

Suppose you wish to forecast how much U.S. real GDP will grow next year. You consult several experts, each of whom is uncertain about the outcome and communicates this uncertainty through lower and upper forecasts. In the fourth-quarter 2018 Duke CFO survey, for example, respondents were asked for ``worst-case'' and ``best-case'' GDP growth estimates associated with a ten-percent chance in each tail~\citep{duke2018questionnaire} (Figure~\ref{fig:survey-question}). The two tail estimates define a central prediction interval with nominal coverage $\alpha=0.8$ (the 10th and 90th percentiles). If only these interval reports are available, how should we combine them into a single point forecast? And would asking for narrower or wider intervals lead to a more reliable aggregate? These questions are particularly relevant to online platforms that elicit and aggregate quantile forecasts.

Quantile-based summaries, including prediction intervals and medians, are natural ways to communicate subjective uncertainty. They are used in finance~\citep{bendavid2013managerial}, climate science~\citep{bamber2022ice}, public health~\citep{bracher2021evaluating}, and energy forecasting~\citep{hong2016probabilistic}. Behavioral evidence provides an additional motivation for studying quantiles directly. As reviewed by \citet{garthwaite2005statistical}, experimental evidence on skewed samples suggests that participants' estimates of sample means were biased toward the corresponding sample medians, whereas their estimates of sample medians were relatively accurate.

However, quantile reports might not contain enough information to yield a point forecast that outperforms even the best constant predictor. Three challenges arise. First, quantile reports can obscure information relevant to prediction under squared loss, for which the posterior mean is optimal. For a binary outcome, for example, posterior probabilities of 0.6 and 0.9 both imply a median report of 1, despite corresponding to different posterior means. Second, the information structure—the joint distribution of the outcome and the experts’ private signals—is unknown. Without this knowledge, we can neither apply the Bayesian combiner nor determine whether experts’ information is redundant or complementary, leaving it unclear how their reports should be combined. Third, when eliciting prediction intervals, the choice of coverage level determines which quantiles are reported and thus what information is available for prediction; how to choose this level is itself an open question.

We study the above challenges under a robust aggregation framework. We consider a bounded outcome, normalized to $[0,1]$, and $n$ experts who each observe a private signal. Each expert truthfully reports a central posterior prediction interval at a common nominal coverage level $\alpha\in(0,1)$. The aggregator observes these reports and knows the reporting format, but not the underlying information structure. Its task is to produce a point forecast evaluated by squared loss. We measure its performance by \emph{regret} relative to an omniscient Bayesian aggregator who knows the information structure and observes all private signals, and therefore predicts the full-signal posterior mean. We seek an aggregation rule that minimizes the worst-case regret over a class of possible information structures, following the robust forecast aggregation framework~\citep{arieli2018robust,neyman2022smarter}.

Existing work on robust forecast aggregation has primarily considered posterior probability reports~\citep{arieli2018robust,levy2022combining,guo2025algorithmic} or posterior mean reports~\citep{neyman2022smarter}. However, these results do not directly determine how to aggregate prediction intervals or medians, nor do they answer which interval coverage level should be elicited. 

As in these prior settings, obtaining a nontrivial worst-case guarantee requires restrictions on the information structure. When arbitrary dependence among private signals is allowed, no aggregation rule based on quantile reports can achieve lower worst-case regret than the constant forecast $1/2$. This is because the general case allows the XOR construction (Example~\ref{ex:xor}) which makes each expert’s posterior distribution given their own signal—and hence every quantile report—uninformative even though the signals jointly reveal the outcome.

We focus on information structures satisfying \emph{weak substitutes}, a diminishing-returns condition on the value of information introduced by \citet{chen2016informational} and used in forecast aggregation by \citet{neyman2022smarter}. Informally, this condition implies that the improvement in expected squared loss from observing an additional signal cannot increase as more signals become available. This assumption rules out the pure XOR example while permitting experts' information to overlap. Under weak substitutes, we derive explicit expressions for the optimal aggregators of interval and median reports, and identify the best interval coverage levels at which the optimal worst-case regret achieves its minimum.

\subsection{Summary of Results}\label{sec:summary-results}

\input{figures/regret_comparison}
\input{figures/reporting_levels}

Our main findings are as follows:
\begin{itemize}
    \item \textbf{Optimal aggregation of prediction intervals.}
    We characterize an optimal rule, which we call \emph{regularized simple average} for prediction intervals (Theorem~\ref{thm:interval-optimality}). It combines the average of the reported interval midpoints with the midpoint of the outcome range, placing a weight on the reports that depends only on the number of experts $n$ and the nominal coverage level $\alpha$.
    Figure~\ref{fig:regret-comparison} compares its worst-case regret with simple averaging of interval midpoints (Corollary~\ref{cor:interval-sa}) and with the optimal worst-case regret attainable from posterior-mean reports (Proposition~\ref{prop:mean-sa}).

    \item \textbf{A phase transition and an optimal interval reporting level.}
    We show that regularized simple average strictly improves upon the optimal constant forecast in worst-case regret if $\alpha<1/\sqrt{n}$, whereas the constant forecast is robust optimal when $\alpha\geq 1/\sqrt{n}$. Moreover, the optimal worst-case regret is minimized at a unique interior coverage level $\alpha^*(n)$, characterized by $n(\alpha^*(n))^2(3+2\alpha^*(n))=1$, strictly below the phase threshold. Both lower and higher reporting levels yield worse guarantees. In particular, for bounded outcomes, the eighty-percent coverage level used in the motivating survey lies in the constant-optimal regime for any $n\geq 2$. Figure~\ref{fig:reporting-levels} illustrates the threshold and the optimal reporting level as the number of experts varies (Theorem~\ref{thm:optimal-level}).
    
    \item \textbf{Optimal aggregation of medians.}
    For median reports, regularized simple average for medians is also optimal. It combines the average reported median with the midpoint of the outcome range, assigning weight $1/n$ to the reported average and the remaining weight to the constant midpoint. This rule strictly improves upon the optimal constant forecast in worst-case regret for every finite $n\geq 2$. Its optimal weight and worst-case regret agree with the corresponding limits of the interval results as $\alpha\downarrow0$, although central prediction intervals need not shrink to a single point when the posterior has multiple medians (Theorem~\ref{thm:median}). Figure~\ref{fig:regret-comparison} compares its worst-case regret with that of simple averaging of medians (Corollary~\ref{cor:median-sa}).
\end{itemize}

Together, these results show that robust aggregation involves not only \emph{how} to combine experts' quantile reports, but also \emph{which} quantiles to ask them to report.

%% file: figures/duke_questionnaire.tex
\begin{figure*}[t]
\centering
\begin{tikzpicture}
\node[anchor=south west, inner sep=0] (questionnaire) at (0,0)
 {\includegraphics[width=0.86\textwidth]{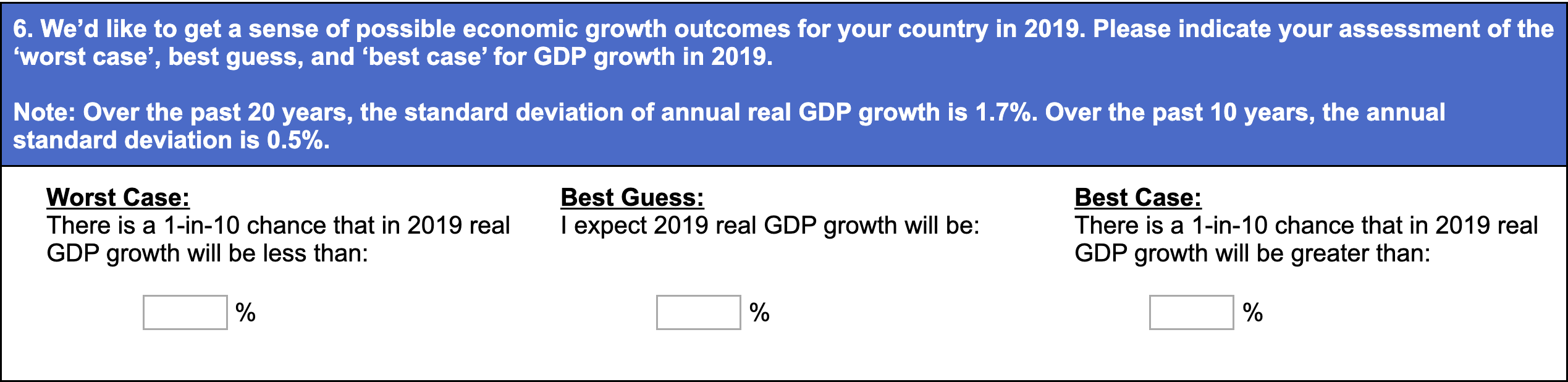}};
\begin{scope}[x={(questionnaire.south east)}, y={(questionnaire.north west)}]
\draw[red!85!black, line width=1pt]
 (0.157,0.408) ellipse [x radius=0.061, y radius=0.052];
\draw[red!85!black, line width=1pt]
 (0.813,0.408) ellipse [x radius=0.061, y radius=0.052];
\node[font=\footnotesize\bfseries, text=red!85!black]
 at (0.167,0.070) {10th percentile};
\node[font=\footnotesize\bfseries, text=red!85!black]
 at (0.823,0.070) {90th percentile};
\end{scope}
\end{tikzpicture}
\caption{The GDP-growth question from the fourth-quarter 2018 Duke CFO survey~\citep{duke2018questionnaire}. The highlighted ten-percent tail probabilities define the 10th and 90th percentiles, forming a central eighty-percent prediction interval.}
\Description{The original GDP-growth survey question has three empty input fields. Added red circles mark the one-in-ten chance in each tail, and red labels identify the lower and upper forecasts as the 10th and 90th percentiles. The middle field asks for a best guess.}
\label{fig:survey-question}
\end{figure*}

%% file: figures/regret_comparison.tex
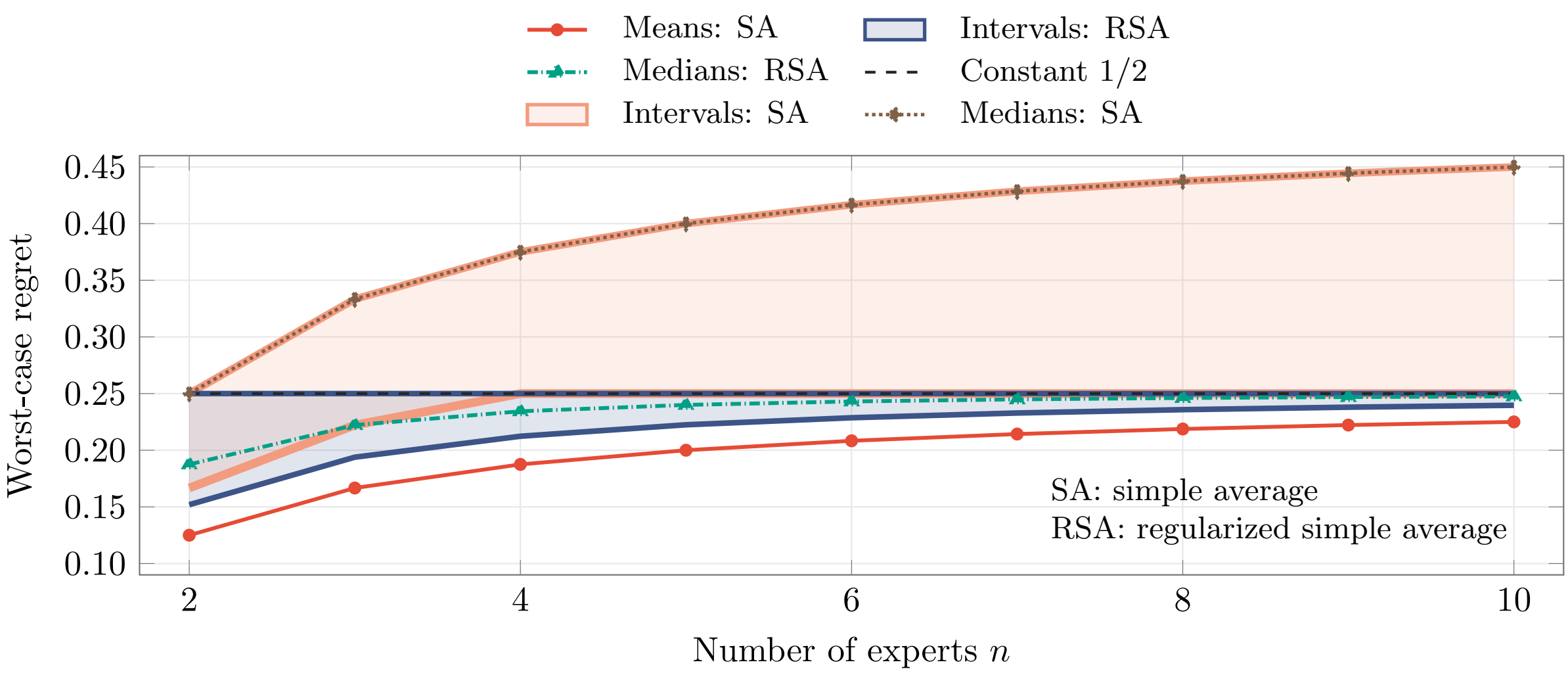
\begin{figure}[t]
\centering
\begin{tikzpicture}
\begin{axis}[
 comparison axis,
 ymin=0.09, ymax=0.46, ytick={0.10,0.15,0.20,0.25,0.30,0.35,0.40,0.45},
 yticklabels={0.10,0.15,0.20,0.25,0.30,0.35,0.40,0.45},
 ylabel={Worst-case regret},
 legend columns=2, legend style={at={(0.5,1.03)}, anchor=south},
]
\addplot[name path=optlo, draw=none, forget plot, samples at={2,...,10}]
 {0.25-2*(cos(acos(2/x-1)/3)-0.5)^3};
\addplot[name path=opthi, draw=none, forget plot, samples at={2,...,10}]
 {0.25};
\addplot[fill=intervalbandblue, fill opacity=0.15, draw=none, forget plot]
 fill between[of=optlo and opthi];
\addplot[name path=midlo, draw=none, forget plot, samples at={2,...,10}]
 {min(0.25,(x-1)/(3*x))};
\addplot[name path=midhi, draw=none, forget plot, samples at={2,...,10}]
 {(x-1)/(2*x)};
\addplot[fill=midpointorange, fill opacity=0.15, draw=none, forget plot]
 fill between[of=midlo and midhi];
\addplot[midpointorange, solid, line width=2.5pt, forget plot,
 samples at={2,...,10}] {min(0.25,(x-1)/(3*x))};
\addplot[midpointorange, solid, line width=2.2pt, forget plot,
 samples at={2,...,10}] {(x-1)/(2*x)};
\addplot[intervalbandblue, solid, line width=1.6pt, forget plot,
 samples at={2,...,10}] {0.25-2*(cos(acos(2/x-1)/3)-0.5)^3};
\addplot[intervalbandblue, solid, line width=1.6pt, forget plot,
 samples at={2,...,10}] {0.25};
\addplot[meanred, line width=1.1pt, mark=*, mark size=1.3pt,
 forget plot,
 samples at={2,...,10}] {(x-1)/(4*x)};
\addplot[mediangreen, densely dashdotted, line width=1.1pt,
 mark=triangle*, mark size=1.7pt, forget plot, samples at={2,...,10}] {0.25-1/(4*x^2)};
\addplot[black!85, dashed, line width=0.8pt, forget plot,
 samples at={2,...,10}] {0.25};
\addplot[rawmedianbrown, densely dotted, line width=1pt,
 mark=diamond*, mark size=1.8pt, forget plot, samples at={2,...,10}] {(x-1)/(2*x)};
\addlegendimage{meanred, line width=1.1pt, mark=*, mark size=1.3pt}
\addlegendentry{Means: SA}
\addlegendimage{area legend, draw=intervalbandblue, fill=intervalbandblue!15, line width=1pt}
\addlegendentry{Intervals: RSA}
\addlegendimage{mediangreen, densely dashdotted, line width=1.1pt, mark=triangle*}
\addlegendentry{Medians: RSA}
\addlegendimage{black!85, dashed, line width=0.8pt}
\addlegendentry{Constant $1/2$}
\addlegendimage{area legend, draw=midpointorange, fill=midpointorange!15, line width=1pt}
\addlegendentry{Intervals: SA}
\addlegendimage{rawmedianbrown, densely dotted, line width=1pt, mark=diamond*}
\addlegendentry{Medians: SA}
\node[anchor=south east, align=left, font=\footnotesize]
  at (rel axis cs:0.97,0.04) {SA: simple average\\RSA: regularized simple average};
\end{axis}
\end{tikzpicture}
\caption{Worst-case regret under weak substitutes. SA is robust optimal for posterior-mean reports, and RSA for intervals and medians. Interval bands span the infimum and supremum over $0<\alpha<1$; SA averages interval midpoints. Median SA coincides with the upper orange boundary. The constant $1/2$ has worst-case regret $0.25$.}
\Description{Worst-case regret plotted against each integer expert count from 2 to 10. The plot compares robust-optimal simple averaging of posterior means, robust-optimal regularized simple average for medians, simple averaging of medians, and the constant one-half baseline. Blue and orange bands show the ranges over reporting levels between zero and one for optimal interval aggregation and simple averaging of interval midpoints, respectively. The constant baseline has worst-case regret one-quarter. The median simple-average curve coincides with the upper orange boundary.}
\label{fig:regret-comparison}
\end{figure}

%% file: figures/reporting_levels.tex
\begin{figure}[t]
\centering
\begin{tikzpicture}
\begin{axis}[
 comparison axis,
 ymin=0, ymax=1, ytick={0,0.2,0.4,0.6,0.8,1},
 ylabel={Reporting level $\alpha$},
 legend columns=2, legend style={at={(0.5,1.03)}, anchor=south},
]
\addplot[name path=phaseboundary, draw=none, forget plot,
 samples at={2,...,10}] {1/sqrt(x)};
\addplot[name path=phasetop, draw=none, forget plot]
 coordinates {(2,1) (10,1)};
\addplot[name path=phasebottom, draw=none, forget plot]
 coordinates {(2,0) (10,0)};
\addplot[fill=phasepositive, fill opacity=0.16, draw=none, forget plot]
 fill between[of=phaseboundary and phasebottom];
\addplot[fill=phasenegative, fill opacity=0.16, draw=none, forget plot]
 fill between[of=phasetop and phaseboundary];
\addplot[intervalbandblue, line width=1.4pt, mark=*, mark size=1.4pt,
 samples at={2,...,10}]
 {1/sqrt(x)};
\addlegendentry{Threshold $\alpha_{\mathrm{thre}}(n)$}
\addplot[optimalred, dashed, line width=1.8pt,
 mark=triangle*, mark size=2pt, mark options={solid, fill=optimalred},
 samples at={2,...,10}]
 {cos(acos(2/x-1)/3)-0.5};
\addlegendentry{Optimal level $\alpha^*(n)$}
\node[font=\small\bfseries, align=center, text=phasenegative!55!black]
  at (axis cs:6.5,0.80) {Constant rule is \textbf{\textcolor{phasepositive!80!black}{OPTIMAL}}};
\node[font=\small\bfseries, align=center, text=phasepositive!45!black]
  at (axis cs:6.5,0.105) {Constant rule is \textbf{\textcolor{phasenegative!80!black}{SUB-OPTIMAL}}};
\end{axis}
\end{tikzpicture}
\caption{The phase threshold $\alpha_{\mathrm{thre}}(n)=1/\sqrt{n}$ and the regret-minimizing interval reporting level $\alpha^*(n)$ both decrease with $n$. Pale blue allows strict improvement over the constant rule in worst-case regret; pale orange marks levels where the constant is robust optimal, including the threshold. The regret-minimizing level lies strictly between zero and the threshold.}
\Description{The decreasing dark blue phase threshold separates a pale blue region below, where strict improvement over the constant rule is possible, from a pale orange region above, where the constant is robust optimal in worst-case regret. Equality at the threshold belongs to the latter region. The coral red regret-minimizing reporting level lies strictly below the phase threshold for every displayed expert count from 2 to 10.}
\label{fig:reporting-levels}
\end{figure}
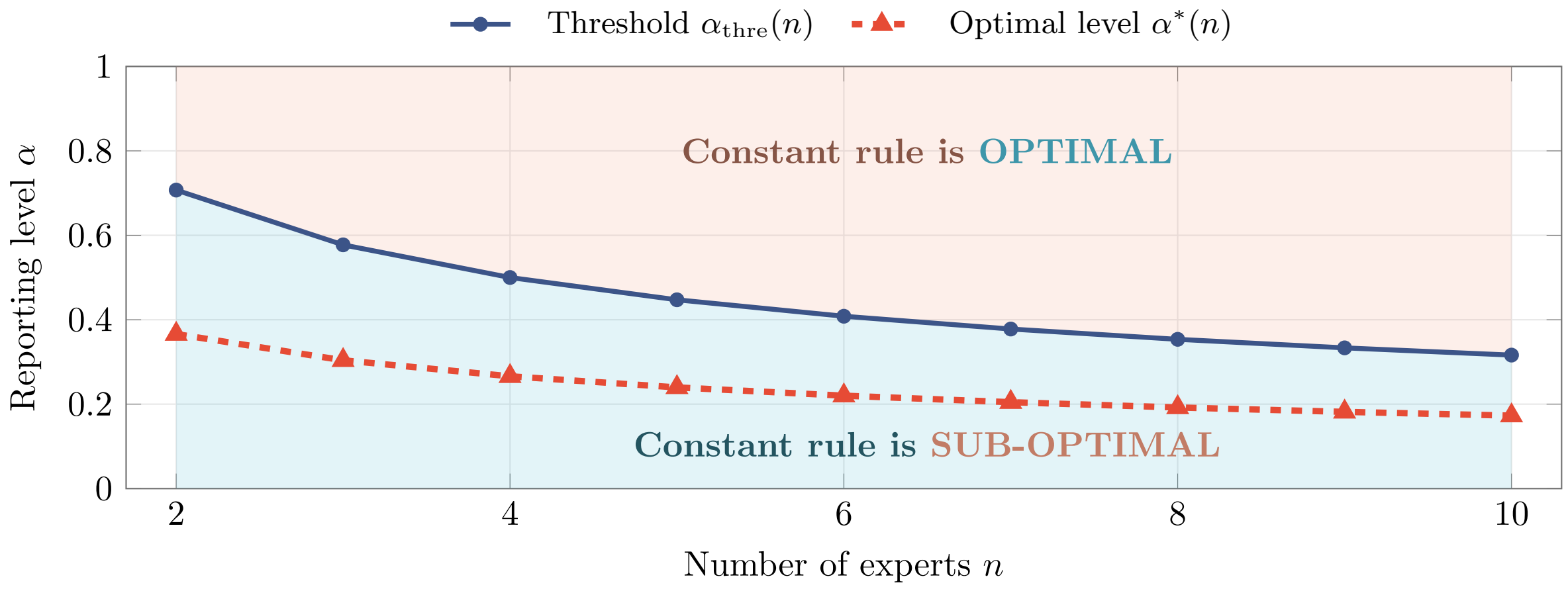

%% file: sections/related_work.tex
\section{Related Work}\label{sec:related-work}

\textbf{Aggregation of quantile statistics.} Quantile-based summaries, including prediction intervals and medians, are natural ways to communicate subjective uncertainty. Proper quantile and interval scoring rules provide truthful mechanisms for eliciting such statistics~\citep{grushka2017quantile,bracher2021evaluating}. Recent studies adopt such mechanisms and investigate how to aggregate the elicited quantile statistics and interval forecasts~\citep{ugander2015wisdom,grushka2017quantile}. We focus on aggregation in this paper, and for simplicity, assume truthful reports without specifying an elicitation mechanism.

Given the experts' quantile reports, existing studies either aggregate them into a point estimate~\citep{ugander2015wisdom,lyon2015collective}, or aggregate them into distributional summaries represented by quantiles, intervals, or full distributions~\citep{genest1992vincentization,lichtendahl2013better,hora2013median,grushka2017quantile,taylor1998combining,giacomini2005evaluation,gaba2017combining}. Regarding the form of aggregation rules, existing studies combine reported values or corresponding quantiles through weighted average~\citep{ugander2015wisdom,lyon2015collective,genest1992vincentization,grushka2017quantile} or use linear combinations of quantile forecasts with coefficients estimated from historical data~\citep{taylor1998combining,giacomini2005evaluation}. Other approaches pool predictive distributions through pointwise averages or medians of cumulative distribution functions~\citep{lichtendahl2013better,hora2013median,gaba2017combining} or logarithmic pooling of densities~\citep{busetti2017quantile}. We characterize a robust optimal point prediction rule, \emph{regularized simple average}, that shrinks the average of interval midpoints or reported medians toward the fixed midpoint of the outcome range, with a weight given explicitly in terms of the number of experts and, for interval reports, the nominal coverage level.


For theoretical guarantees, existing theoretical work characterizes when quantile aggregation improves on random selection under proper scoring rules~\citep{grushka2017quantile} and assesses gains from combining forecasts through encompassing tests~\citep{giacomini2005evaluation}. Other work derives maximum-likelihood aggregation rules under specific signal models~\citep{ugander2015wisdom}. In this paper, we characterize the robust optimal aggregation rule with minimum worst-case regret relative to the omniscient aggregator under weak substitutes~\citep{chen2016informational,neyman2022smarter}, optimizing over \emph{all} possible aggregation rules.

\textbf{Robust Aggregation Frameworks.} The robust aggregation framework, introduced by~\citet{arieli2018robust}, seeks rules with worst-case performance guarantees over a specified class of information structures. In their work, they consider a binary outcome scenario where two Bayesian experts truthfully report their posterior beliefs.

Subsequent work varies the assumptions on information structures and the information available to the aggregator. Several studies assume that experts' signals are conditionally independent~\citep{arieli2020identifiable,guo2025algorithmic,chen2026prior}. Other approaches study unknown dependence with known individual information structures~\citep{arieli2025random,deoliveira2026robust,levy2022combining}. \citet{neyman2022smarter} study posterior-mean reports for general real-valued outcomes under \emph{weak substitutes} and \emph{projective substitutes}.

Another line of work varies the information elicited from experts. Second-order reports can improve aggregation~\citep{pan2024robust,kong2024peer}, while \citet{frongillo2025queries} develop a general framework of elicitable queries and relate query complexity to aggregation accuracy. To our knowledge, our work is the first to characterize robust optimal aggregation of prediction intervals and medians for general bounded outcomes under squared loss and weak substitutes.

%% file: sections/model.tex
\section{Model}
\label{sec:model}

Fix an integer $n\ge2$ and a nominal coverage level $\alpha\in(0,1)$. Write $\beta=(1-\alpha)/2$ and $[n]=\{1,\ldots,n\}$. The outcome is a random variable $Y$ taking values in $\mathcal Y=[0,1]$. Each expert $i$ observes a private signal $S_i$ taking values in $\mathcal S_i$. We use lowercase letters $y$ and $s_i$ for their realizations, and write $S=(S_1,\ldots,S_n)$ and $s=(s_1,\ldots,s_n)$. An \emph{information structure} is a joint distribution $\theta\in\Delta(\mathcal Y\times\mathcal S_1\times\cdots\times\mathcal S_n)$ of the outcome and signals. Expert $i$ forms a Bayesian posterior and truthfully reports the central prediction interval
\begin{equation}\label{eq:reports}
 \begin{aligned}
 Z_i=[L_i,U_i]
 &=\bigl[Q_{\theta,i}(\beta\mid S_i),Q_{\theta,i}(1-\beta\mid S_i)\bigr],\\
 Q_{\theta,i}(p\mid s_i)
 &=\inf\{y\in[0,1]:\PP_\theta(Y\le y\mid S_i=s_i)\ge p\}.
 \end{aligned}
\end{equation}
We use this lower generalized quantile convention throughout. The two quantile levels assign the same nominal tail probability $\beta$ on each side. The resulting interval has posterior coverage at least \(\alpha\), with possibly larger coverage when the posterior has atoms. A realized interval report is denoted by $z_i=[\ell_i,u_i]$.

For median reports, we set $Z_i=Q_{\theta,i}(1/2\mid S_i)$, using the same symbol $Z_i$ for the random report and $z_i$ for its realization. This replaces the two quantile levels by their common limit $1/2$ as $\alpha\downarrow0$. Figure~\ref{fig:posterior-quantiles} illustrates both reporting formats.

\input{figures/posterior_quantiles}

The aggregator knows $n$ and the reporting format, including $\alpha$ for interval reports, but not the information structure $\theta$ or the private signals. After observing the report vector $Z=(Z_1,\ldots,Z_n)$, it gives a single point prediction of the outcome. Let \(\mathcal Z\) denote the space of a single expert's report under the chosen reporting format. An aggregation rule is a measurable map
$f:\mathcal Z^n\to[0,1]$, with output $f(z)$ at a realized report vector $z=(z_1,\ldots,z_n)$.

\paragraph{Benchmark.}
We evaluate predictions using squared loss. An \emph{omniscient aggregator} knows both $\theta$ and all private signals, so its optimal prediction is the full-signal posterior mean
\[
 M_\theta=\E_\theta[Y\mid S],\qquad
 m_\theta(s)=\E_\theta[Y\mid S=s].
\]
Given aggregation rule $f$, its regret relative to this benchmark is
\[
 R(f,\theta)=\E_\theta[(f(Z)-Y)^2]-\E_\theta[(M_\theta-Y)^2].
\]

We seek a rule minimizing worst-case regret, which we call a \emph{robust optimal} rule. For interval reports with nominal coverage level $\alpha$, define the optimal worst-case regret over a class $\Theta$ of information structures by
\begin{equation}
\label{eq:minimax-definition}
    V_\Theta(n,\alpha)=\inf_{f:\mathcal Z^n\to[0,1]}\sup_{\theta\in\Theta}R(f,\theta).
\end{equation}
If $\Theta$ is unrestricted, no aggregation rule has lower worst-case regret than the constant rule $f\equiv1/2$, as the following example shows.

\begin{example}[XOR information structure]\label{ex:xor}
Let $(S_1,S_2)$ be uniformly distributed on $\{0,1\}^2$ and set $Y=S_1\oplus S_2$, where $\oplus$ denotes addition modulo two. For $n>2$, let the remaining signals be constant.

Every expert's posterior is Bernoulli$(1/2)$, so all interval reports are $[0,1]$ at every $0<\alpha<1$; all lower median reports are $0$. Thus the report vector is constant, while the full signals reveal $Y$ exactly. If a rule outputs $h$ at this report vector, then
\[
 R(f,\theta)=\tfrac12h^2+\tfrac12(1-h)^2
 =\tfrac14+(h-\tfrac12)^2\ge\tfrac14.
\]
The constant rule $f\equiv1/2$ satisfies $R(f,\theta)=\E_\theta[(M_\theta-1/2)^2]\le1/4$ for every information structure. Together with the XOR lower bound, this shows that the optimal worst-case regret over unrestricted information structures is $1/4$ for either reporting format.
\end{example}

In the XOR example, the first two signals are \emph{information complements}: neither is informative alone, but together they determine the outcome. To exclude such scenarios, we introduce the \emph{weak substitutes} condition \citep{chen2016informational,neyman2022smarter}.

\begin{definition}[Weak substitutes]\label{def:ws}
For $A\subseteq[n]$, write $S_A=(S_i)_{i\in A}$ and define the information value of this coalition by
\[
 v_\theta(A)=\E_\theta[(Y-\E_\theta Y)^2]
             -\E_\theta[(Y-\E_\theta[Y\mid S_A])^2].
\]
This is the improvement in expected squared loss from observing $S_A$, relative to the uninformed prediction $\E_\theta Y$; in particular, $v_\theta(\varnothing)=0$. An information structure satisfies weak substitutes if
\[
v_\theta(A\cup\{i\})-v_\theta(A)\ge v_\theta(B\cup\{i\})-v_\theta(B)
\]
for every $A\subseteq B\subseteq[n]$ and $i\notin B$.
\end{definition}

Weak substitutes means that the ex ante marginal value of an additional signal is non-increasing as the observed coalition grows. Figure~\ref{fig:weak-substitutes} illustrates this property in a symmetric information structure, plotting the information value of a coalition's signals against its size. The following equivalent variance representation justifies this measure of information value and will be used throughout the paper.

\input{figures/weak_substitutes}

\begin{proposition}[Variance representation of weak substitutes]\label{prop:ws-variance}
For $A\subseteq[n]$, let $M_{\theta,A}=\E_\theta[Y\mid S_A]$, so $M_{\theta,\varnothing}=\E_\theta Y$ and $M_{\theta,[n]}=M_\theta$. Then $v_\theta(A)=\Var_\theta(M_{\theta,A})$. Consequently, the weak substitutes condition is equivalent to
\[
 \begin{aligned}
 &\Var_\theta(M_{\theta,A\cup\{i\}})-\Var_\theta(M_{\theta,A})\ge\Var_\theta(M_{\theta,B\cup\{i\}})-\Var_\theta(M_{\theta,B})
 \end{aligned}
\]
for every $A\subseteq B\subseteq[n]$ and $i\notin B$. In particular, it implies
\begin{equation}\label{eq:ws-sum}
 \sum_{i=1}^n\Var_\theta(M_{\theta,\{i\}})
 \ge\Var_\theta(M_\theta).
\end{equation}
\end{proposition}
The proof is given in Appendix~\ref{app:ws-variance}. Let $\Theta_{\WS}$ be the class of information structures satisfying weak substitutes, and abbreviate $V_{\Theta_{\WS}}(n,\alpha)$ as $V(n,\alpha)$. When $\theta$ is fixed, we write $M_i=M_{\theta,\{i\}}$ and $M=M_\theta$ for the individual and full-signal posterior means.

%% file: figures/posterior_quantiles.tex
\begin{figure}[!htbp]
\centering
\begin{tikzpicture}[x=6.8cm,y=1.15cm,font=\footnotesize]
\def\lowerquantile{0.2123171283}
\def\medianquantile{0.3857275681}
\def\upperquantile{0.5824535745}
\path[fill=midpointorange!30]
  (0,0) -- plot[domain=0:\lowerquantile,samples=35]
  (\x,{12*\x*(1-\x)^2}) -- (\lowerquantile,0) -- cycle;
\path[fill=intervalbandblue!20]
  (\lowerquantile,0) -- plot[domain=\lowerquantile:\upperquantile,samples=45]
  (\x,{12*\x*(1-\x)^2}) -- (\upperquantile,0) -- cycle;
\path[fill=midpointorange!30]
  (\upperquantile,0) -- plot[domain=\upperquantile:1,samples=45]
  (\x,{12*\x*(1-\x)^2}) -- (1,0) -- cycle;
\draw[->,black!65] (0,0) -- (1.055,0) node[right] {$y$};
\draw[->,black!65] (0,0) -- (0,2.05);
\draw[intervalbandblue,thick]
  plot[domain=0:1,samples=100] (\x,{12*\x*(1-\x)^2});
\draw[intervalbandblue,densely dashed]
  (\lowerquantile,0) -- (\lowerquantile,{12*\lowerquantile*(1-\lowerquantile)^2});
\draw[intervalbandblue,densely dashed]
  (\upperquantile,0) -- (\upperquantile,{12*\upperquantile*(1-\upperquantile)^2});
\draw[mediangreen,thick,densely dashed]
  (\medianquantile,0) -- (\medianquantile,{12*\medianquantile*(1-\medianquantile)^2});
\node[anchor=west] at (0.025,2.03) {Posterior density given private signal $s_i$};
\node at (0.12,0.42) {$\beta$};
\node[intervalbandblue] at (0.46,0.64) {$\alpha$};
\node at (0.73,0.28) {$\beta$};
\node[below=2pt] at (0,0) {$0$};
\node[below=2pt] at (1,0) {$1$};
\node[below=2pt,intervalbandblue] at (\lowerquantile,0) {$\ell_i$};
\node[below=2pt,mediangreen] at (\medianquantile,0) {median};
\node[below=2pt,intervalbandblue] at (\upperquantile,0) {$u_i$};
\draw[intervalbandblue,thick]
  (\lowerquantile,-0.45) -- (\lowerquantile,-0.55)
  -- (\upperquantile,-0.55) -- (\upperquantile,-0.45);
\node[below=2pt,intervalbandblue] at (0.3974,-0.55) {Interval report $[\ell_i,u_i]$};
\end{tikzpicture}
\caption{Quantile reports summarize an expert's posterior belief. For this continuous example, $[\ell_i,u_i]=\big[Q_{\theta,i}(\beta\mid S_i=s_i),Q_{\theta,i}(1-\beta\mid S_i=s_i)\big]$ covers the central $\alpha$ probability, and the dashed green line is the median $Q_{\theta,i}(1/2\mid S_i=s_i)$. With atoms, the interval's coverage can exceed $\alpha$.}
\Description{A skewed posterior density on zero to one. Two blue dashed lines mark the lower and upper reported quantiles. The blue area between them has probability alpha, and the orange tails each have probability beta. A green dashed line marks the median inside the reported interval.}
\label{fig:posterior-quantiles}
\end{figure}
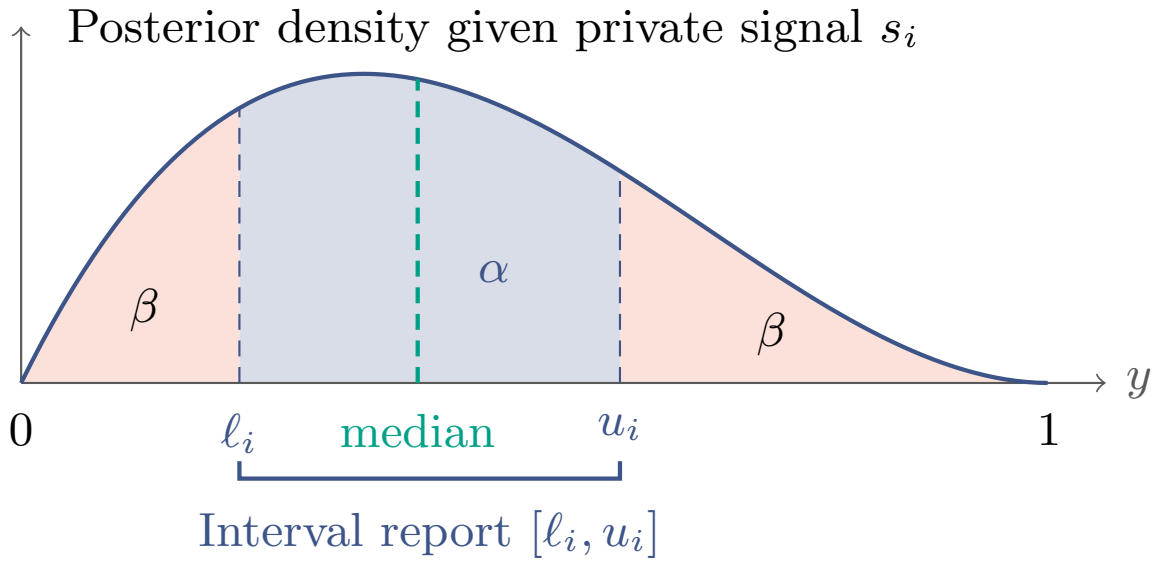

%% file: figures/weak_substitutes.tex
\begin{figure}[!htbp]
\centering
\begin{tikzpicture}[font=\footnotesize]
\begin{axis}[
 width=\linewidth, height=4cm,
 xmin=0, xmax=5.25, ymin=0, ymax=0.3,
 xtick={0,1,2,3,4,5}, ytick={0,0.125,0.25},
 yticklabels={$0$,$1/8$,$1/4$},
 xlabel={Coalition size $|A|$},
 ylabel={$\Var_\theta(\E_\theta[Y\mid S_A])$},
 tick label style={font=\footnotesize},
 label style={font=\footnotesize},
 axis lines=left, axis line style={black!55},
 tick style={black!55}, grid=major, grid style={black!8},
]
\draw[black!45,densely dashed] (axis cs:0,0.25) -- (axis cs:5,0.25);
\node[anchor=south east,black!60] at (axis cs:5,0.25) {$\Var_\theta(Y)$};
\draw[midpointorange,densely dashed] (axis cs:0,0) -- (axis cs:1,0);
\draw[midpointorange,thick,<->] (axis cs:1,0) -- (axis cs:1,0.125);
\node[anchor=west,rawmedianbrown] at (axis cs:1.12,0.0625) {First signal};
\draw[midpointorange,densely dashed] (axis cs:2,0.1875) -- (axis cs:3,0.1875);
\draw[midpointorange,thick,<->] (axis cs:3,0.1875) -- (axis cs:3,0.21875);
\node[anchor=west,rawmedianbrown] at (axis cs:3.12,0.198) {Third signal};
\addplot[intervalbandblue,thick,mark=*,mark size=1.7pt,
 domain=0:5,samples=6] {(1-2^(-x))/4};
\end{axis}
\end{tikzpicture}
\caption{Weak substitutes in a symmetric example: each signal independently reveals a fair binary outcome with probability $1/2$ and otherwise reveals nothing, giving $v_\theta(A)=(1-2^{-|A|})/4$.}
\Description{A plot of posterior-mean variance against coalition size. Blue points at integer sizes zero through five are joined by an increasing curve with decreasing slopes, approaching the prior variance of one quarter. Orange vertical segments show that the first signal adds more explained variance than the third. The example is symmetric, so coalition value depends only on its size.}
\label{fig:weak-substitutes}
\end{figure}
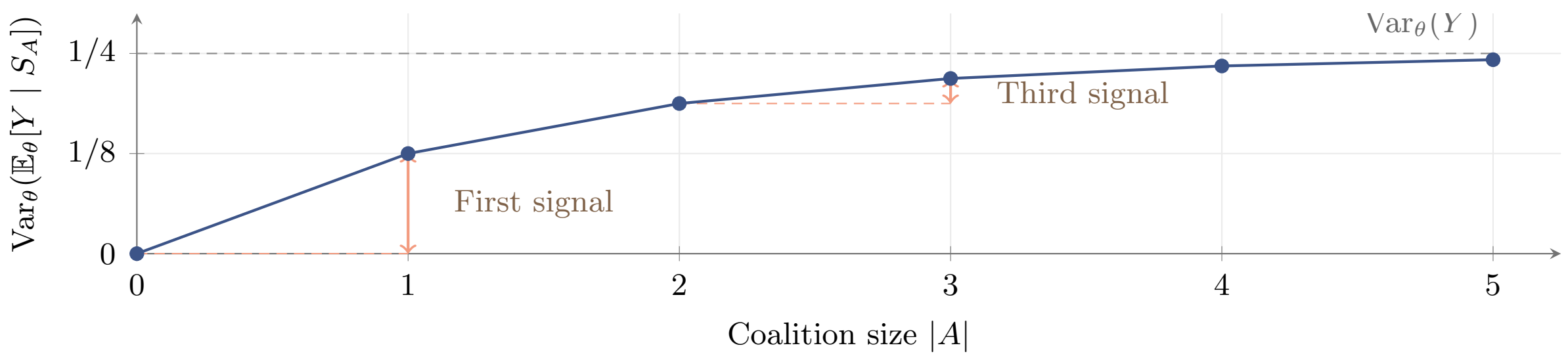

%% file: sections/aggregation_of_intervals.tex
\section{Aggregation of Prediction Intervals}\label{sec:interval-aggregation}

For interval reports $z_i=[\ell_i,u_i]$, write $c(z_i)=(\ell_i+u_i)/2$ for each reported midpoint and $\bar c(z)=n^{-1}\sum_i c(z_i)$ for their simple average.

\begin{theorem}[Regularized Simple Average for Prediction Intervals]\label{thm:interval-optimality}
For every $n\ge2$ and $0<\alpha<1$, a robust optimal aggregation rule for $\Theta=\Theta_{\WS}$ in Equation~\eqref{eq:minimax-definition} is
\begin{equation}\label{eq:optimal-rule}
 \begin{aligned}
 f^*_{n,\alpha}(z)
 &=\eta^*_{n,\alpha}\bar c(z)
       +(1-\eta^*_{n,\alpha})\frac12\\
 &=\frac12+\eta^*_{n,\alpha}\left(\bar c(z)-\frac12\right).
 \end{aligned}
\end{equation}
Let $a_n=(\sqrt{1+8/n}-1)/4$, so that $na_n(1+2a_n)=1$. The weight is
\begin{equation}\label{eq:optimal-weight}
 \eta^*_{n,\alpha}=
 \begin{cases}
 \displaystyle\frac{1+n\alpha}{n(1+\alpha)},&0<\alpha\le a_n,\\[5pt]
 2\alpha,&a_n<\alpha<\min\{1/2,1/\sqrt n\},\\[2pt]
 1,&1/2\le\alpha<1/\sqrt n,\\[2pt]
 0,&1/\sqrt n\le\alpha<1.
 \end{cases}
\end{equation}
We call this rule \emph{regularized simple average for prediction intervals}. Its worst-case regret equals $V(n,\alpha)$, where
\begin{equation}\label{eq:interval-value}
 V(n,\alpha)=
 \begin{cases}
 \displaystyle\frac{(n-1)(n+1+2n\alpha)}{4n^2(1+\alpha)^2},&0<\alpha\le a_n,\\[6pt]
 \displaystyle\frac{(n-1)\alpha}{n(1+\alpha)}
                  +\left(\frac12-\alpha\right)^2,
       &a_n<\alpha<\min\{1/2,1/\sqrt n\},\\[6pt]
 \displaystyle\frac{n-1}{4n(1-\alpha^2)},
       &1/2\le\alpha<1/\sqrt n,\\[5pt]
 \displaystyle\frac14,&1/\sqrt n\le\alpha<1.
 \end{cases}
\end{equation}
The third branch in both formulas applies only when $n=2$ or $3$. In particular, $V(n,\alpha)<1/4$ if and only if $\alpha<1/\sqrt n$.
\end{theorem}

The rule moves the average reported midpoint toward the center $1/2$, placing weight $\eta^*_{n,\alpha}$ on the reports and the remaining weight on $1/2$. This adjustment accounts for the fact that an interval midpoint need not equal the expert's posterior mean. At the threshold $\alpha_{\mathrm{thre}}(n)=1/\sqrt n$, every weight in $[0,\min\{2/\sqrt n,1\}]$ is optimal, and in Equation~\eqref{eq:optimal-weight} we select the constant rule. The adjacent formulas for worst-case regret $V(n,\alpha)$ agree at every boundary.

\paragraph{Optimal reporting level.}
Before proving Theorem~\ref{thm:interval-optimality} and explaining the weights $\eta^{*}_{n,\alpha}$, we use the formulas to identify the nominal coverage level $\alpha^*(n)$ that minimizes the worst-case regret.

\begin{theorem}[Optimal reporting level]\label{thm:optimal-level}
For each integer $n\ge2$, the optimal worst-case regret $V(n,\alpha)$ has a unique minimizer over $0<\alpha<1$, denoted by $\alpha^*(n)$. This minimizer satisfies
\begin{equation}\label{eq:optimal-level-root}
 \begin{aligned}
 n\bigl(\alpha^*(n)\bigr)^2\bigl(3+2\alpha^*(n)\bigr)&=1,\\
 a_n<\alpha^*(n)&<\min\{1/2,1/\sqrt n\},
 \end{aligned}
\end{equation}
and strictly decreases as $n$ increases, with $\alpha^*(n)\sim(3n)^{-1/2}$ as $n\to\infty$. Moreover,
\begin{equation*}
 \begin{aligned}
 \inf_{0<\alpha<1}V(n,\alpha)&=\frac14-2\bigl(\alpha^*(n)\bigr)^3=\frac14-\frac{2+o(1)}{3\sqrt3\,n^{3/2}},\\
 \sup_{0<\alpha<1}V(n,\alpha)&=\frac14.
 \end{aligned}
\end{equation*}
The infimum is attained only at $\alpha=\alpha^*(n)$, and the supremum is attained exactly when $1/\sqrt n\le\alpha<1$.
\end{theorem}

\begin{proof}
The second branch of Equation~\eqref{eq:interval-value} has derivative
\[
 \frac{n\alpha^2(3+2\alpha)-1}{n(1+\alpha)^2}.
\]
Its numerator is strictly increasing and changes sign from negative to positive between $a_n$ and $\min\{1/2,1/\sqrt n\}$. The first branch strictly decreases and the third (when present) strictly increases, so continuity gives the unique minimizer. The supremum follows directly from Theorem~\ref{thm:interval-optimality}. Substituting Equation~\eqref{eq:optimal-level-root} gives the minimum value. Finally, $\alpha^2(3+2\alpha)$ is strictly increasing, so its solution at $1/n$ strictly decreases with $n$. Equation~\eqref{eq:optimal-level-root} gives $\alpha^*(n)\to0$ and $\sqrt n\,\alpha^*(n)\to1/\sqrt3$, yielding the stated asymptotics.
\end{proof}

Low reporting levels can pool weak and strong evidence into the same extreme report, whereas high levels can pool beliefs on opposite sides of $1/2$ into the same wide interval. The optimal reporting level balances these two losses of information. The minimum worst-case regret \(\inf_\alpha V(n,\alpha)\) increases toward \(1/4\) as \(n\) grows. Thus, more expert reports yield a larger worst-case regret relative to the corresponding omniscient aggregator benchmark.

\paragraph{Simple-average comparisons.}
Taking $\eta=1$ gives simple averaging of interval midpoints, $f_1(z)=\bar c(z)$. Its exact worst-case regret determines the orange band in Figure~\ref{fig:regret-comparison}.

\begin{corollary}[Simple average for prediction intervals]\label{cor:interval-sa}
For every $n\ge2$ and $0<\alpha<1$, write $V_{\mathrm{mid}}(n,\alpha)=\sup_{\theta\in\Theta_{\WS}}R(f_1,\theta)$ for the worst-case regret achieved by the simple average. Then
\begin{equation}\label{eq:midpoint-regret}
 V_{\mathrm{mid}}(n,\alpha)=
 \begin{cases}
 \displaystyle\min\left\{\frac{1-\alpha}{2},
              \frac{n-1}{2n(1+\alpha)}\right\},
       &0<\alpha\le\frac12,\\[6pt]
 \displaystyle\min\left\{\frac14,
              \frac{n-1}{4n(1-\alpha^2)}\right\},
       &\frac12<\alpha<1.
 \end{cases}
\end{equation}
Consequently,
\begin{equation}\label{eq:midpoint-envelope}
 \begin{aligned}
 \inf_{0<\alpha<1}V_{\mathrm{mid}}(n,\alpha)
   &=\min\left\{\frac14,\frac{n-1}{3n}\right\},\\
 \sup_{0<\alpha<1}V_{\mathrm{mid}}(n,\alpha)
   &=\frac{n-1}{2n}.
 \end{aligned}
\end{equation}
\end{corollary}
Appendix~\ref{app:interval-sa} proves the corollary using a method similar to that for Theorem~\ref{thm:interval-optimality}. Under posterior mean reports, simple average is robust optimal with worst-case regret $(n-1)/(4n)$. The argument of \citet{neyman2022smarter} implies this optimality result in our bounded-outcome setting. Appendix~\ref{app:mean-sa} gives a direct proof.

We prove Theorem~\ref{thm:interval-optimality} in three steps. Section~\ref{sec:binary-upper} considers binary outcomes, deriving the upper bound and explaining the weights in Equation~\eqref{eq:optimal-weight}. Section~\ref{sec:interval-extension} extends this upper bound to all bounded outcomes in $[0,1]$ through Lemma~\ref{lem:binary-to-bounded}, which also applies to median reports. Section~\ref{sec:binary-lower} then gives matching lower bounds using binary information structures, completing the proof.

\subsection{Upper Bound for Binary Outcomes}\label{sec:binary-upper}

For now suppose $Y\in\{0,1\}$. For a weight $\eta\in[0,1]$, we can write the resulting regularized simple average as
\[
  f_\eta(z)=\frac1n\sum_{i=1}^n q_\eta(z_i),
  \qquad q_\eta(z_i)=\eta c(z_i)+(1-\eta)\frac12,
\]
where $q_\eta(z_i)$ denotes the contribution of expert $i$'s report. We also write \(q_\eta(\ell,u)\) for \(q_\eta([\ell,u])\).

For binary outcomes, each expert must hold a Bernoulli posterior. When expert $i$'s posterior mean is $m_i=\E_{\theta}[Y\mid S_i=s_i]$, using $\beta=(1-\alpha)/2$ and the lower-quantile convention in Equation~\eqref{eq:reports} we have:
\begin{equation}\label{eq:binary-report-regions}
\begin{array}{c|c|c}
 \text{posterior mean } m_i & \text{report } z_i & q_\eta(z_i)\\ \hline
 0\le m_i\le\beta &[0,0]&(1-\eta)/2\\
 \beta<m_i\le1-\beta &[0,1]&1/2\\
 1-\beta<m_i\le1 &[1,1]&(1+\eta)/2.
\end{array}
\end{equation}

Fix an information structure $\theta\in\Theta_{\WS}$. Write $\mu=\E_\theta Y$, $v_i=\Var_\theta(M_i)$, and $v_N=\Var_\theta(M)$ for the prior mean and the variances of the individual and full-signal posterior means. Jensen's inequality and $\E_{\theta}[M\mid S_i]=M_i$ give
\begin{equation}\label{eq:binary-decomposition}
 \begin{aligned}
 R(f_\eta,\theta)&=\E_{\theta}[(M-f_\eta(Z))^2] \\
 &\le\frac1n\sum_i\E_{\theta}[(M-q_\eta(Z_i))^2]\\
 &=v_N-\frac1n\sum_i v_i
       +\frac1n\sum_i\E_{\theta}[(M_i-q_\eta(Z_i))^2].
 \end{aligned}
\end{equation}
The weak substitutes condition implies $\sum_i v_i\ge v_N$ by Proposition~\ref{prop:ws-variance}, while $v_N\le\Var(Y)=\mu(1-\mu)$. To control the remaining reporting error, we bound it in terms of the posterior variance $m_i(1-m_i)$:
\begin{equation}\label{eq:binary-certificate}
 (m_i-q_\eta(z_i))^2\le(\gamma-1)m_i(1-m_i)+\frac{(1-\eta)^2}{4}.
\end{equation}
The last term comes from degenerate Bernoulli posteriors $m_i=0$ or $1$.

\begin{lemma}\label{lem:binary-coefficient}
For $0<\alpha<1$ and $0\le\eta\le1$, inequality~\eqref{eq:binary-certificate} holds for every Bernoulli posterior with mean $m_i$ and its corresponding contribution $q_{\eta}(z_i)$ if and only if
\[
 \gamma\ge\Gamma_\alpha(\eta)
 :=\max\left\{\frac{2\eta}{1+\alpha},
              \frac{\eta(2-\eta)}{1-\alpha^2}\right\}.
\]
\end{lemma}

The two terms come from the extreme-report and central-report regions, respectively. Appendix~\ref{app:binary-certificate} proves the lemma. For any valid $\gamma\in[0,n]$, substituting inequality~\eqref{eq:binary-certificate} into inequality~\eqref{eq:binary-decomposition} gives
\[
 \begin{aligned}
 R(f_\eta,\theta)
 &\le v_N-\frac\gamma n\sum_i v_i
       +(\gamma-1)\mu(1-\mu)+\frac{(1-\eta)^2}{4}\\
 &\le\frac{n-1}{n}\gamma\mu(1-\mu)+\frac{(1-\eta)^2}{4}\\
 &\le\frac{n-1}{4n}\gamma+\frac{(1-\eta)^2}{4}.
 \end{aligned}
\]
Here $\E[M_i(1-M_i)]=\mu(1-\mu)-v_i$; the second inequality uses both $\sum_i v_i\ge v_N$ and $v_N\le\mu(1-\mu)$, with $1-\gamma/n\ge0$.

For $\alpha\le1/\sqrt n$, the coefficient $\Gamma_\alpha(\eta)$ lies in $[0,n]$ for every $\eta\in[0,1]$. Choosing $\gamma=\Gamma_\alpha(\eta)$ yields
\begin{equation}\label{eq:weight-objective}
 \begin{aligned}
 R(f_\eta,\theta)&\le\mathcal B_{n,\alpha}(\eta)\\
 &:=\frac{(1-\eta)^2}{4}+\frac{n-1}{4n}
 \max\left\{\frac{2\eta}{1+\alpha},
              \frac{\eta(2-\eta)}{1-\alpha^2}\right\}.
 \end{aligned}
\end{equation}
The two requirements coincide at $\eta=2\alpha$, apart from $\eta=0$, so
\begin{equation}\label{eq:weight-objective-pieces}
 \mathcal B_{n,\alpha}(\eta)=
 \begin{cases}
 \displaystyle\frac14-
 \frac{1-n\alpha^2}{4n(1-\alpha^2)}\eta(2-\eta),
       &0\le\eta\le\min\{2\alpha,1\},\\[6pt]
 \displaystyle\frac{(n-1)\eta}{2n(1+\alpha)}+\frac{(1-\eta)^2}{4},
       &2\alpha\le\eta\le1.
 \end{cases}
\end{equation}

For $\alpha<1/\sqrt n$, the first piece decreases with $\eta$, and the second is a quadratic minimized at $(1+n\alpha)/(n(1+\alpha))$. Minimizing these two pieces gives the positive weights in Equation~\eqref{eq:optimal-weight}.
At small levels the quadratic minimum determines the weight. It meets $2\alpha$ when $n\alpha(1+2\alpha)=1$, giving the transition $a_n$. At intermediate levels the best weight lies where the two constraints meet; if $2\alpha$ reaches one before the phase threshold, the best weight is one. These are the three positive-weight cases in Equation~\eqref{eq:optimal-weight}. 

When $\alpha=1/\sqrt{n}$, both branches in $\mathcal B_{n,\alpha}(\eta)$ are no less than $1/4$, indicating that this upper bound no longer yields a strict improvement over the constant rule. The matching lower bound in Section~\ref{sec:binary-lower} shows that no aggregation rule can achieve worst-case regret below \(1/4\) at or above this threshold. This explains the phase threshold $\alpha_{\mathrm{thre}}(n)=1/\sqrt n$.

\begin{proposition}[Binary upper bound]\label{prop:binary-upper}
For every $n\ge2$ and $0<\alpha<1$, regularized simple average for prediction intervals satisfies
\[
 R(f^*_{n,\alpha},\theta)\le V(n,\alpha)
\]
for every $\theta\in\Theta_{\WS}$ with $Y\in\{0,1\}$, where the right-hand side is the expression in Equation~\eqref{eq:interval-value}.
\end{proposition}

\begin{proof}
For $\alpha<1/\sqrt n$, substitute the three positive weights in Equation~\eqref{eq:optimal-weight} into inequality~\eqref{eq:weight-objective}. They give the first three branches of Equation~\eqref{eq:interval-value}, respectively. For $\alpha\ge1/\sqrt n$, the proposed rule is $f^*_{n,\alpha}\equiv1/2$, whose regret is $\E[(M-1/2)^2]\le1/4$.
\end{proof}

\subsection{Extension to Bounded Outcomes}\label{sec:interval-extension}

The core idea is that we can decompose the general posterior over $[0,1]$ into a family of binary random variables. Specifically, given expert $i$'s posterior $W_i\sim\PP(Y\in\cdot\mid S_i=s_i)$, we can construct
\[
B_{i,t}=\mathbf1\{W_i>t\}
\]
for every threshold $t\in(0,1)$. Moreover, we have a similar decomposition for the quantiles:
\begin{align*}
   \ell_{i,t}=\mathbf1\{\ell_i>t\}, &\quad u_{i,t}=\mathbf1\{u_i>t\}.
\end{align*}
For each threshold $t$, $B_{i,t}$ is a binary random variable, and the thresholded endpoints $\ell_{i,t}$ and $u_{i,t}$ are the lower quantiles of $B_{i,t}$ at levels $\beta$ and $1-\beta$, respectively. Then we have
\begin{equation}
\label{eq:interval-threshold-representation}
    \begin{aligned}
        W_i&=\int_0^1B_{i,t}\,dt, \\
        q_\eta(\ell_i,u_i)&=\int_0^1q_\eta(\ell_{i,t},u_{i,t})\,dt.
    \end{aligned}
\end{equation}
The following lemma transfers a binary posterior inequality to arbitrary posterior distributions supported on $[0,1]$ through such representations.

\begin{lemma}[From binary posterior bounds to regret]\label{lem:binary-to-bounded}
Fix $n\ge2$, $0\le\gamma\le n$, and $\kappa\ge0$. Let $W$ take values in $[0,1]$, set $B_t=\mathbf1\{W>t\}$, and suppose a measurable family of predictions $h_t\in[0,1]$ satisfies $h=\int_0^1h_t\,dt$ and, for almost every $t\in(0,1)$,
\begin{equation}\label{eq:threshold-certificate}
 (\E B_t-h_t)^2\le(\gamma-1)\Var(B_t)+\kappa.
\end{equation}
Then
\begin{equation}\label{eq:shared-posterior-bound}
 \begin{aligned}
 (\E W-h)^2&\le(\gamma-1)\Var(W)\\
 &\quad+\frac{(n-1)\gamma}{n}\E[W(1-W)]+\kappa.
 \end{aligned}
\end{equation}
For any $\theta\in\Theta_{\WS}$, suppose these hypotheses hold for each expert's conditional posterior $W_i$ and almost every private-signal realization, with report-based prediction $h=q_i(z_i)$ and the same $\gamma$ and $\kappa$. Then the aggregation rule $f(z)=n^{-1}\sum_i q_i(z_i)$ satisfies
\begin{equation}\label{eq:shared-regret-bound}
 R(f,\theta)\le\frac{(n-1)\gamma}{4n}+\kappa.
\end{equation}
\end{lemma}

The proof is given in Appendix~\ref{app:binary-to-bounded}.

\begin{proposition}[Upper bound for interval aggregation]\label{prop:interval-upper}
For every $n\ge2$, $0<\alpha\le1/\sqrt n$, $\theta\in\Theta_{\WS}$ with bounded outcome $Y\in[0,1]$, and $\eta\in[0,1]$,
\begin{equation}\label{eq:interval-upper}
 \begin{aligned}
 R(f_\eta,\theta)
 &\le\frac{(n-1)\Gamma_\alpha(\eta)}{4n}+\frac{(1-\eta)^2}{4}=\mathcal B_{n,\alpha}(\eta).
 \end{aligned}
\end{equation}
Consequently, regularized simple average for prediction intervals satisfies $R(f^*_{n,\alpha},\theta)\le V(n,\alpha)$ for every $0<\alpha<1$, where the right-hand side is the expression in Equation~\eqref{eq:interval-value}.
\end{proposition}
For interval reports, Lemma~\ref{lem:binary-coefficient} supplies inequality~\eqref{eq:threshold-certificate} with $h_t=q_\eta(\ell_{i,t},u_{i,t})$, $\gamma=\Gamma_\alpha(\eta)$, and $\kappa=(1-\eta)^2/4$. When $0<\alpha\le1/\sqrt n$, we have $\gamma\in[0,n]$, so Lemma~\ref{lem:binary-to-bounded} yields upper bound $\mathcal B_{n,\alpha}(\eta)$. Above this threshold, the selected rule is constant and its regret is at most $1/4$. Appendix~\ref{app:interval-upper} gives the detailed proof.

\subsection{Matching Lower Bound}\label{sec:binary-lower}

\begin{theorem}[Secret sharing with a common posterior]\label{thm:interval-lower}
For every $n\ge2$, $0<\alpha<1$, and $\varepsilon>0$, there is a finite information structure $\theta\in\Theta_{\WS}$ with a binary outcome such that every aggregation rule $f$ satisfies
\[
 R(f,\theta)\ge V(n,\alpha)-\varepsilon,
\]
where the right-hand side uses the expression in Equation~\eqref{eq:interval-value}. If $\alpha>1/\sqrt n$, one such structure satisfies $R(f,\theta)\ge1/4$ for every $f$.
\end{theorem}

The construction builds on Shamir's secret-sharing scheme, where sufficiently many private signals recover the outcome while fewer shares reveal nothing~\citep{shamir1979share}. \citet{neyman2022smarter} randomize the recovery threshold uniformly to satisfy the weak substitutes condition and obtain the lower bound $(n-1)/(4n)$. In that construction, interval reports introduce no additional information loss relative to posterior means, so the bound falls strictly below $V(n,\alpha)$.

To create that additional ambiguity and obtain a matching lower bound, we add an additional common signal $U$. Each expert observes $U$, the recovery threshold $K$, and a private share. This construction ensures that each expert's posterior is $\mathrm{Bernoulli}(U)$, while all signals together reveal $Y$. We choose the distribution of $U$ to hide as much information as possible within each report category. Conditional on $U$ and $K$, the private shares can exhibit threshold complementarities, while the ex ante coalition information values satisfy weak substitutes. The aggregator cannot distinguish different posterior means that induce the same reports, indicating a lower bound on the regret relative to the omniscient benchmark which predicts $Y$ exactly in this construction. By choosing the distribution of \(U\) appropriately, we can match every branch of Equation~\eqref{eq:interval-value}. Appendix~\ref{app:interval-lower} gives the full construction.

Combining Theorem~\ref{thm:interval-lower} with Proposition~\ref{prop:interval-upper} proves Theorem~\ref{thm:interval-optimality}. Appendix~\ref{app:interval-proof} gives the complete argument.

%% file: sections/aggregation_of_medians.tex
\section{Aggregation of Medians}\label{sec:median}

For median reports $z=(z_1,\ldots,z_n)$, write $\bar z=n^{-1}\sum_{i=1}^n z_i$ for their simple average. Denote the robust optimal worst-case regret by
\begin{equation*}
 V_{\mathrm{med}}(n)
 =\inf_{f:[0,1]^n\to[0,1]}
       \sup_{\theta\in\Theta_{\WS}}R(f,\theta).
\end{equation*}

\begin{theorem}[Regularized Simple Average for Medians]\label{thm:median}
For every $n\ge2$, a robust optimal aggregation rule for median reports under weak substitutes is
\begin{equation}\label{eq:median-rule}
 \begin{aligned}
 f^*_{n,\mathrm{med}}(z)
 &=\eta^*_{n,\mathrm{med}}\bar z
       +(1-\eta^*_{n,\mathrm{med}})\frac12
       \\
 &=\frac12+\frac1n\left(\bar z-\frac12\right),
 \end{aligned}
\end{equation}
where $\eta^*_{n,\mathrm{med}}=\frac1n$. We call this rule \emph{regularized simple average for medians}. Its worst-case regret is
\begin{equation}\label{eq:median-value}
 \sup_{\theta\in\Theta_{\WS}}R(f^*_{n,\mathrm{med}},\theta)
 =V_{\mathrm{med}}(n)=\frac14-\frac1{4n^2}.
\end{equation}
The same rule and worst-case regret apply under any method for selecting a median based on each expert's signal, including when the worst case also ranges over such selections.
\end{theorem}

As for prediction intervals, regularized simple average combines the average report with the center $1/2$. The rule improves on the optimal constant prediction in worst-case regret by $1/(4n^2)$ for every finite $n\ge2$.

\paragraph{Relation to prediction intervals.}
The optimal weight and worst-case regret agree with the corresponding limits in Theorem~\ref{thm:interval-optimality}:
\begin{equation*}
 \eta^*_{n,\mathrm{med}}=\lim_{\alpha\downarrow0}\eta^*_{n,\alpha}
 =\frac1n,\qquad
 V_{\mathrm{med}}(n)=\lim_{\alpha\downarrow0}V(n,\alpha).
\end{equation*}
Nevertheless, taking the limit $\alpha\downarrow0$ in the interval theorem does not by itself prove the median result. For a fair Bernoulli posterior, every positive-level central interval is $[0,1]$, whereas the lower generalized median is $0$. If every expert has this posterior, regularized simple average for prediction intervals always outputs $1/2$, while regularized simple average for medians outputs $(n-1)/(2n)$. We therefore establish the median result directly.

\paragraph{Simple-average comparisons.}
Taking $\eta=1$ gives simple averaging of medians, $f_{\mathrm{med},1}(z)=\bar z$. Its exact worst-case regret determines the median SA curve in Figure~\ref{fig:regret-comparison}, which coincides with the upper interval-SA boundary in Equation~\eqref{eq:midpoint-envelope}.

\begin{corollary}[Simple averaging for medians]\label{cor:median-sa}
For every $n\ge2$, simple averaging $f_{\mathrm{med},1}(z)=\bar z$ has exact worst-case regret
\[
 \sup_{\theta\in\Theta_{\WS}}R(f_{\mathrm{med},1},\theta)
 =\frac{n-1}{2n},
\]
under any median-selection method.
\end{corollary}
Appendix~\ref{app:median-sa} proves the corollary.

The proof of Theorem~\ref{thm:median} follows the same structure as the interval analysis in Section~\ref{sec:interval-aggregation}. Section~\ref{sec:median-upper} first establishes a binary posterior inequality analogous to inequality~\eqref{eq:binary-certificate}, then applies Lemma~\ref{lem:binary-to-bounded} to extend it to bounded outcomes and derive the upper bound. Section~\ref{sec:median-lower} obtains a matching lower bound by choosing a different distribution for the common signal $U$ in the secret-sharing construction of Section~\ref{sec:binary-lower}. Section~\ref{sec:median-proof} combines the two bounds to prove Theorem~\ref{thm:median}.

\subsection{Upper Bound}\label{sec:median-upper}

Fix $n\ge2$ and $0\le\eta\le1$. Define the contribution of expert $i$'s median report and the resulting aggregation rule by
\[
 \begin{aligned}
 q_{\mathrm{med},\eta}(z_i)&=\eta z_i+\frac{1-\eta}{2},\\
 f_{\mathrm{med},\eta}(z)&=\frac1n\sum_{i=1}^n q_{\mathrm{med},\eta}(z_i).
 \end{aligned}
\]

\paragraph{Binary posterior inequality.}
First focus on binary outcomes. For a Bernoulli posterior with mean $m_i$, the unique median is zero if $m_i<1/2$ and one if $m_i>1/2$; when $m_i=1/2$, every $z_i\in[0,1]$ is a median. For every such median report, we have
\begin{equation}\label{eq:median-binary-certificate}
 (m_i-q_{\mathrm{med},\eta}(z_i))^2
 \le(2\eta-1)m_i(1-m_i)+\frac{(1-\eta)^2}{4}.
\end{equation}
This is the median counterpart of inequality~\eqref{eq:binary-certificate} with $\gamma=2\eta$.

\paragraph{Extension to bounded outcomes.}
Lemma~\ref{lem:binary-to-bounded} transfers inequality~\eqref{eq:median-binary-certificate} to bounded outcomes with $\gamma=2\eta$ and $\kappa=(1-\eta)^2/4$.

\begin{proposition}[Upper bound for median aggregation]\label{prop:median-upper}
For every $\theta\in\Theta_{\WS}$ with bounded outcome $Y\in[0,1]$ and every $\eta\in[0,1]$,
\begin{equation}\label{eq:median-upper}
 \begin{aligned}
 R(f_{\mathrm{med},\eta},\theta)
 &\le\frac{\eta(n-1)}{2n}+\frac{(1-\eta)^2}{4}\\
 &=\frac14-\frac1{4n^2}
       +\frac14\left(\eta-\frac1n\right)^2.
 \end{aligned}
\end{equation}
This bound holds for any median selection method.
\end{proposition}

\begin{proof}
Fix $\theta\in\Theta_{\WS}$. For each expert's posterior $W_i\sim\PP(Y\in\cdot\mid S_i=s_i)$ and any reported median $z_i$, define
\[
 B_{i,t}=\mathbf1\{W_i>t\},\qquad z_{i,t}=\mathbf1\{z_i>t\}.
\]
If $z_i>t$, then $\PP(B_{i,t}=1)\ge\PP(W_i\ge z_i)\ge1/2$; if $z_i\le t$, then $\PP(B_{i,t}=0)\ge\PP(W_i\le z_i)\ge1/2$. Hence $z_{i,t}$ is a median of $B_{i,t}$, including when that binary posterior has multiple medians. Threshold integration and the affine form of $q_{\mathrm{med},\eta}$ give
\[
 q_{\mathrm{med},\eta}(z_i)=\int_0^1q_{\mathrm{med},\eta}(z_{i,t})\,dt.
\]
Applying Lemma~\ref{lem:binary-to-bounded} with inequality~\eqref{eq:median-binary-certificate}, $h_t=q_{\mathrm{med},\eta}(z_{i,t})$, $\gamma=2\eta\in[0,n]$, and $\kappa=(1-\eta)^2/4$ gives inequality~\eqref{eq:median-upper}. The argument applies to any signal-dependent median selection, so the bound holds for every such method.
\end{proof}

\subsection{Matching Lower Bound}\label{sec:median-lower}

\begin{proposition}[Lower bound for median aggregation]\label{prop:median-lower}
For every $n\ge2$ and $\varepsilon>0$, there is a finite binary information structure $\theta\in\Theta_{\WS}$ with unique posterior medians such that every aggregation rule $f:[0,1]^n\to[0,1]$ satisfies
\[
 R(f,\theta)\ge\frac14-\frac1{4n^2}-\varepsilon.
\]
\end{proposition}

\begin{proof}
We use the common-posterior construction of Section~\ref{sec:binary-lower}, choosing the following law of $U$ to hide different posterior probabilities behind identical median reports.

Fix $0<\delta<1/2$, and put $\rho=(n-1)/n$ and $b=1/2-\delta$. Define
\[
 \PP(U=0)=\PP(U=1)=\frac{1-\rho}{2},
 \qquad \PP(U=b)=\PP(U=1-b)=\frac{\rho}{2}.
\]
This common posterior mean $U$ has expectation $1/2$ and variance
\[
 \Var(U)=\frac14-\rho b(1-b)
     =\frac1{4n}+\frac{n-1}{n}\delta^2\ge\frac1{4n}.
\]
Applying the common-posterior construction of Appendix~\ref{app:common-posterior} to this law yields a finite information structure $\theta_\delta$ with $M_i=U$ for every expert and $M=Y$. Since $\Var(U)\ge1/(4n)$, Lemma~\ref{lem:common-posterior-ws} ensures that $\theta_\delta\in\Theta_{\WS}$.

Every singleton posterior has a unique median: zero when \(U\in\{0,b\}\) and one when \(U\in\{1-b,1\}\). Consequently, the only report vectors are $0^n$ and $1^n$, each with probability $1/2$. Denote $h_0=f(0^n)$ and $h_1=f(1^n)$ for an arbitrary aggregation rule $f$. Since $M=Y$, the regret satisfies
\[
 \begin{aligned}
 R(f,\theta_\delta)
 &=\rho b(1-\rho b)+\frac12(h_0-\rho b)^2+\frac12\bigl(h_1-(1-\rho b)\bigr)^2\\
 &\ge \rho b(1-\rho b)\longrightarrow\frac14-\frac1{4n^2}
 \qquad\text{as }\delta\downarrow0.
 \end{aligned}
\]
The lower bound is independent of $f$. For every $\varepsilon>0$, sufficiently small $\delta$ therefore gives the required finite structure in $\Theta_{\WS}$ simultaneously for all rules. Its unique medians make the construction independent of the median selection method.
\end{proof}

\subsection{Proof of Theorem~\ref{thm:median}}\label{sec:median-proof}

\begin{proof}
Choosing $\eta=1/n$ in Proposition~\ref{prop:median-upper} gives the rule in Equation~\eqref{eq:median-rule} and the upper bound
\[
 \sup_{\theta\in\Theta_{\WS}}R(f^*_{n,\mathrm{med}},\theta)
 \le\frac14-\frac1{4n^2}.
\]
The rule lies in $[0,1]$ by convexity. Proposition~\ref{prop:median-lower} implies that every rule has worst-case regret at least $1/4-1/(4n^2)-\varepsilon$ for every $\varepsilon>0$. Letting $\varepsilon\downarrow0$ and taking the infimum over rules yields the matching lower bound for $V_{\mathrm{med}}(n)$. Together these bounds prove Equation~\eqref{eq:median-value} and optimality of the regularized simple average in Equation~\eqref{eq:median-rule}. The upper bound allows any signal-dependent median selection, while the lower-bound structures have unique medians. Thus the result holds for each median selection method and when the worst case also ranges over such methods.
\end{proof}

%% file: sections/discussion.tex
\section{Discussion and Future Work}\label{sec:discussion}

We characterize robust optimal rules for interval and median reports and show how interval coverage affects optimal worst-case regret.

There are several future directions. First, we assume truthful reports and central prediction intervals at a common nominal coverage level. Future work could study asymmetric intervals or heterogeneous reporting levels across experts. Second, we mainly focus on the weak substitutes condition. Considering more general information structures could provide a richer representation of the aggregator’s knowledge. Finally, the proposed rules have not been evaluated empirically. Robust optimality concerns worst-case regret over the class of information structures, while other rules may achieve lower regret in particular structures. Experiments with real-world data could evaluate how the proposed rules compare with alternatives in practice.

%% file: appendices/weak_substitutes.tex
\section{Proof of Proposition~\ref{prop:ws-variance}}\label{app:ws-variance}

\begin{proof}
Fix $\theta$ and write $\mu=\E_\theta Y$. For every $A\subseteq[n]$, the tower property gives $\E_\theta M_{\theta,A}=\mu$. Since $M_{\theta,A}-\mu$ is determined by $S_A$ and $\E_\theta[Y-M_{\theta,A}\mid S_A]=0$, we have
\[
 \begin{aligned}
 &\E_\theta[(Y-M_{\theta,A})(M_{\theta,A}-\mu)]=\E_\theta\!\left[(M_{\theta,A}-\mu)\E_\theta[Y-M_{\theta,A}\mid S_A]\right]=0.
 \end{aligned}
\]
Expanding $Y-\mu=(Y-M_{\theta,A})+(M_{\theta,A}-\mu)$ therefore yields
\[
 \begin{aligned}
 \E_\theta[(Y-\mu)^2]
 &=\E_\theta[(Y-M_{\theta,A})^2]+\E_\theta[(M_{\theta,A}-\mu)^2].
 \end{aligned}
\]
By the definition of $v_\theta(A)$ and  $\E_\theta M_{\theta,A}=\mu$, it follows that
\[
 v_\theta(A)=\E_\theta[(M_{\theta,A}-\mu)^2]
            =\Var_\theta(M_{\theta,A}).
\]

To obtain inequality~\eqref{eq:ws-sum}, let $A_i=\{1,\ldots,i\}$ for $i=1,\ldots,n$ and $A_0=\varnothing$. Apply weak substitutes with $\varnothing\subseteq A_{i-1}$ and $i\notin A_{i-1}$:
\[
 v_\theta(A_i)-v_\theta(A_{i-1})
 \le v_\theta(\{i\})-v_\theta(\varnothing)=v_\theta(\{i\}).
\]
Summing these inequalities and using $v_\theta(\varnothing)=0$, we obtain
\[
 \begin{aligned}
 v_\theta([n])
 &=\sum_{i=1}^n\bigl(v_\theta(A_i)-v_\theta(A_{i-1})\bigr)\le\sum_{i=1}^n v_\theta(\{i\}).
 \end{aligned}
\]
The variance representation now gives inequality~\eqref{eq:ws-sum}.
\end{proof}

%% file: appendices/interval_aggregation_proofs.tex
\section{Proofs for Optimal Interval Aggregation}\label{app:interval-aggregation}

\subsection{Proof of Lemma~\ref{lem:binary-coefficient}}\label{app:binary-certificate}

\begin{proof}
Fix expert $i$'s Bernoulli posterior with mean $m_i$ and corresponding interval report $z_i=[\ell_i,u_i]$. In the low-report region \(0\le m_i\le\beta\), we have $q_\eta(z_i)=(1-\eta)/2$. Subtracting the right-hand side of inequality~\eqref{eq:binary-certificate} from the left-hand side gives
\[
 m_i\bigl[\eta-\gamma(1-m_i)\bigr],\qquad 0\le m_i\le\beta.
\]
This is nonpositive throughout the region exactly when $\gamma\ge\eta/(1-\beta)=2\eta/(1+\alpha)$. Necessity follows by taking $m_i=\beta$. The high-report region $1-\beta<m_i\le 1$ gives the same condition after replacing $m_i$ by $1-m_i$.

For the central report $\beta<m_i\le 1-\beta$, we have $q_\eta(z_i)=1/2$, and the corresponding difference is
\[
 \frac{\eta(2-\eta)}4-\gamma m_i(1-m_i),
 \qquad \beta<m_i\le1-\beta.
\]
The smallest posterior variance in this region is $\beta(1-\beta)=(1-\alpha^2)/4$, attained at $m_i=1-\beta$. Hence the equivalent condition for the central-report region is $\gamma\ge\eta(2-\eta)/(1-\alpha^2)$.

Combining these two conditions gives $\Gamma_\alpha(\eta)$.
\end{proof}

\subsection{Proof of Lemma~\ref{lem:binary-to-bounded}}\label{app:binary-to-bounded}

\begin{proof}
On a fixed probability space, define the residual
\[
 \begin{aligned}
 \mathcal F(W,h)&=(\E W-h)^2-(\gamma-1)\Var(W)-\frac{(n-1)\gamma}{n}\E[W(1-W)]\\
 &=(\E W-h)^2+\left(1-\frac\gamma n\right)\Var(W)+\frac{(n-1)\gamma}{n}\bigl((\E W)^2-\E W\bigr).
 \end{aligned}
\]
The second expression is jointly convex in $(W,h)$ because $0\le\gamma\le n$. Since $W=\int_0^1B_t\,dt$ and $B_t(1-B_t)=0$, inequality~\eqref{eq:threshold-certificate} and Jensen's inequality yield
\[
 \mathcal F(W,h)\le\int_0^1\mathcal F(B_t,h_t)\,dt\le\kappa,
\]
which proves inequality~\eqref{eq:shared-posterior-bound}.

For the regret bound, fix $\theta\in\Theta_{\WS}$ and write $\mu=\E Y$, $\sigma^2=\Var(Y)$, $v_i=\Var(M_i)$, $v_N=\Var(M)$, and $d_Y=\E[Y(1-Y)]$. The decomposition in inequality~\eqref{eq:binary-decomposition} also holds for bounded outcomes and contributions $q_i(Z_i)$:
\[
 \begin{aligned}
 R(f,\theta)&\le v_N-\frac1n\sum_i v_i+\frac1n\sum_i\E[(M_i-q_i(Z_i))^2].
 \end{aligned}
\]
Apply inequality~\eqref{eq:shared-posterior-bound} to each expert's conditional posterior, take expectations over that expert's signal realizations, and then average over the $n$ experts. By the law of total variance, the averaged posterior variance is $\sigma^2-n^{-1}\sum_i v_i$, while the averaged $\E[W_i(1-W_i)]$ term equals $d_Y$. Thus
\[
 \begin{aligned}
 R(f,\theta)&\le v_N-\frac\gamma n\sum_i v_i+(\gamma-1)\sigma^2+\frac{(n-1)\gamma}{n}d_Y+\kappa\\
 &\le\left(1-\frac\gamma n\right)v_N+(\gamma-1)\sigma^2+\frac{(n-1)\gamma}{n}d_Y+\kappa\\
 &\le\frac{(n-1)\gamma}{n}(\sigma^2+d_Y)+\kappa\\
 &\le\frac{(n-1)\gamma}{4n}+\kappa.
 \end{aligned}
\]
The second inequality uses $\sum_i v_i\ge v_N$ from weak substitutes and $\gamma\ge0$. The third uses $v_N\le\sigma^2$ and $\gamma\le n$, and the last uses $\sigma^2+d_Y=\mu(1-\mu)\le1/4$.
\end{proof}

\subsection{Proof of Proposition~\ref{prop:interval-upper}}\label{app:interval-upper}

\begin{proof}
Fix $n\ge2$, $0<\alpha\le1/\sqrt n$, $\theta\in\Theta_{\WS}$, and $\eta\in[0,1]$. Set $\gamma=\Gamma_\alpha(\eta)\in[0,n]$ and $\kappa=(1-\eta)^2/4$. For each expert's posterior, the thresholded endpoints in Section~\ref{sec:interval-extension} are valid binary interval reports, so Lemma~\ref{lem:binary-coefficient} verifies inequality~\eqref{eq:threshold-certificate} with $h_t=q_\eta(\ell_{i,t},u_{i,t})$. Equation~\eqref{eq:interval-threshold-representation} gives the required representation of $h=q_\eta(z_i)$. Lemma~\ref{lem:binary-to-bounded} therefore yields
\[
 R(f_\eta,\theta)\le\frac{(n-1)\Gamma_\alpha(\eta)}{4n}+\frac{(1-\eta)^2}{4},
\]
which is inequality~\eqref{eq:interval-upper}. For $\alpha<1/\sqrt n$, substituting the weights in Equation~\eqref{eq:optimal-weight} gives the first three branches of Equation~\eqref{eq:interval-value}. For $\alpha\ge1/\sqrt n$, the selected rule is constant and has regret $\E[(M-1/2)^2]\le1/4$ because $M\in[0,1]$. Thus the proposed rule has the stated upper bound for every $0<\alpha<1$.
\end{proof}

\subsection{Proof of Theorem~\ref{thm:interval-lower}}\label{app:interval-lower}

We first define a common-posterior information structure and characterize when it satisfies weak substitutes. We then give the specific choice of the distribution of $U$ for the low and intermediate reporting levels and for the midpoint and constant regions, respectively. Combining these constructions proves Theorem~\ref{thm:interval-lower}.

\subsubsection{A common-posterior information structure}\label{app:common-posterior}

We adapt the construction of \citet[Proposition~2.5]{neyman2022smarter} by adding a common signal $U$. Let $U\in[0,1]$ have finite support and $\E U=1/2$, and draw $Y\mid U\sim\operatorname{Bernoulli}(U)$. Independently of $(U,Y)$, draw $K$ uniformly from $\{2,\ldots,n\}$. Fix a prime $q>n$ and, conditional on $K$, draw independent uniform coefficients $A_1,\ldots,A_{K-1}\in\mathbb F_q$, independently of $(U,Y)$. Following Shamir's scheme~\citep{shamir1979share}, define
\[
 P(t)=Y+\sum_{j=1}^{K-1}A_jt^j,
 \qquad S_i=(U,K,P(i)),\quad i\in[n].
\]
Let $\theta$ be the resulting joint distribution of $(Y,S_1,\ldots,S_n)$. Every expert observes $U$ and $K$ together with a private share $P(i)$; the aggregator observes only the reports.

\begin{lemma}[Weak substitutes condition]\label{lem:common-posterior-ws}
In the common-posterior construction above, $Y\sim\operatorname{Bernoulli}(1/2)$, each expert's posterior is $\operatorname{Bernoulli}(U)$, and the full set of signals reveals $Y$. Hence $M_i=U$ for every $i$ and $M=Y$.

For every nonempty $A\subseteq[n]$, its information value is
\begin{equation}\label{eq:common-posterior-values}
 v_\theta(A)=\Var(U)+\frac{|A|-1}{n-1}\left(\frac14-\Var(U)\right),
\end{equation}
with $v_\theta(\varnothing)=0$. Therefore, the information structure $\theta$ defined above satisfies weak substitutes if and only if
\[
 \Var(U)\ge\frac1{4n}.
\]
\end{lemma}

\begin{proof}
Any $K$ distinct shares $(P(i_1),\dots,P(i_K))$ recover $Y=P(0)$ by polynomial interpolation. For fewer than $K$ shares, the map from the uniform coefficients to the evaluations has full row rank, so they provide no further information about $Y$ conditional on $(U,K)$. Hence, for every nonempty coalition $A$ of size $k$,
\[
 M_{A}=
 \begin{cases}
 Y,&K\le k,\\
 U,&K>k.
 \end{cases}
\]
Since $K\ge2$, each expert's posterior is $\operatorname{Bernoulli}(U)$, and since $K\le n$, $M=Y$. Both $U$ and $Y$ have mean $1/2$, and $\Var(Y)=1/4$. Independence of $K$ from $(U,Y)$ and $\PP(K\le k)=(k-1)/(n-1)$ therefore give Equation~\eqref{eq:common-posterior-values} by Proposition~\ref{prop:ws-variance}.

The information value depends only on coalition size. Its first marginal increment is $\Var(U)$, and every subsequent increment is $(1/4-\Var(U))/(n-1)$. Thus the weak substitutes condition holds if and only if
\[
 \Var(U)\ge\frac{1/4-\Var(U)}{n-1}
 \quad\Longleftrightarrow\quad\Var(U)\ge\frac1{4n}.
\]
\end{proof}

\subsubsection{The low and intermediate reporting levels}

\begin{lemma}\label{lem:shrinkage-lower}
If $0<\alpha<\min\{1/2,1/\sqrt n\}$, then for every $\varepsilon>0$ there is a finite information structure $\theta\in\Theta_{\WS}$ with a binary outcome such that $R(f,\theta)\ge V(n,\alpha)-\varepsilon$ for every aggregation rule $f$.
\end{lemma}

\begin{proof}
Set
\[
 \begin{gathered}
 x_n(\alpha)=\frac{n-1}{2n(1+\alpha)},\qquad
 x=\min\{x_n(\alpha),1/2-\alpha\},\\
 p=\frac{1/(4n)-\alpha^2/4}{2(1-\beta)(\beta-x)}.
 \end{gathered}
\]
Since $\beta-x_n(\alpha)=(1-n\alpha^2)/(2n(1+\alpha))>0$, these choices give $0<x<\beta$, $0<p\le1/2$, and
\begin{equation}\label{eq:category-information}
 \frac{\alpha^2}{4}+2p(1-\beta)(\beta-x)=\frac1{4n}.
\end{equation}

For sufficiently small $0<\delta<\min\{\beta,\alpha/2\}$, put $b=\beta-\delta$ and $d=\beta+\delta$, and choose $t_\delta$ to satisfy
\begin{equation}\label{eq:category-calibration}
 2p\,t_\delta b(1-b)+(1-2p)d(1-d)=\frac{n-1}{4n}.
\end{equation}
Equation~\eqref{eq:category-information} implies $t_\delta\to x/\beta\in(0,1)$, so $0<t_\delta<1$ for all sufficiently small $\delta$. Let $C$ have probabilities $p,1-2p,p$ on $\mathrm{low},\mathrm{mid},\mathrm{high}$, respectively, and define
\[
 U\mid C\sim
 \begin{cases}
 0\text{ or }b,\quad\text{with probabilities }1-t_\delta,t_\delta,
       &C=\mathrm{low},\\
 d\text{ or }1-d,\quad\text{with probabilities }1/2,1/2,
       &C=\mathrm{mid},\\
 1\text{ or }1-b,\quad\text{with probabilities }1-t_\delta,t_\delta,
       &C=\mathrm{high}.
 \end{cases}
\]
Zero-probability categories are omitted. Symmetry gives $\E U=1/2$, and Equation~\eqref{eq:category-calibration} gives
\begin{equation*}
 \Var(U)=\frac14-\E[U(1-U)]=\frac1{4n}
 \quad\text{for every such }\delta.
\end{equation*}
Use this law of $U$ in the construction of Appendix~\ref{app:common-posterior}, and denote the resulting structure by $\theta_\delta$. Lemma~\ref{lem:common-posterior-ws} gives $\theta_\delta\in\Theta_{\WS}$, $M_i=U$, $M=Y$, and $v_{\theta_\delta}(A)=|A|/(4n)$.

Since $b<\beta<d<1/2$, every expert reports $[0,0]$, $[0,1]$, or $[1,1]$ according to $C=\mathrm{low},\mathrm{mid},\mathrm{high}$. The conditional outcome means are $x_\delta=t_\delta b$, $1/2$, and $1-x_\delta$, respectively. Since $M=Y$, every aggregation rule satisfies
\begin{equation*}
 R(f,\theta_\delta)\ge
 2p x_\delta(1-x_\delta)+\frac{1-2p}{4}.
\end{equation*}
As $\delta\downarrow0$, $x_\delta\to x$. The limiting lower bound is
\begin{equation*}
 \begin{aligned}
 L(p,x)&=2px(1-x)+\frac{1-2p}{4}\\
 &=
 \begin{cases}
 x_n(\alpha)(1-x_n(\alpha)),&\alpha\le a_n,\\[2pt]
 \displaystyle\frac14-
       \frac{\alpha(1-n\alpha^2)}{n(1+\alpha)},&\alpha>a_n.
 \end{cases}
 \end{aligned}
\end{equation*}
These are the first two branches of Equation~\eqref{eq:interval-value}. For any $\varepsilon>0$, one sufficiently small $\delta$ therefore gives $R(f,\theta_\delta)\ge V(n,\alpha)-\varepsilon$ simultaneously for every $f$.
\end{proof}

\subsubsection{The midpoint and constant regions}

\begin{lemma}\label{lem:central-lower}
For every $n\ge2$, $0<\alpha<1$, and $\varepsilon>0$, there is a finite information structure $\theta\in\Theta_{\WS}$ with a binary outcome such that every aggregation rule $f$ satisfies
\[
 R(f,\theta)\ge\min\left\{\frac14,\frac{n-1}{4n(1-\alpha^2)}\right\}-\varepsilon.
\]
If $\alpha>1/\sqrt n$, one such structure gives $R(f,\theta)\ge1/4$ for every $f$. Consequently,
\[
 V(n,\alpha)\ge\min\left\{\frac14,\frac{n-1}{4n(1-\alpha^2)}\right\}.
\]
\end{lemma}

\begin{proof}
Choose $\beta<\epsilon<1/2$ and define
\begin{equation}\label{eq:four-posterior-law}
 \begin{aligned}
 \PP(U=\epsilon)=\PP(U=1-\epsilon)&=\frac w2,\\
 \PP(U=0)=\PP(U=1)&=\frac{1-w}{2},
 \end{aligned}
\end{equation}
where
\begin{equation*}
 w=\min\left\{1,\frac{n-1}{4n\epsilon(1-\epsilon)}\right\}.
\end{equation*}
Then $\E U=1/2$ and $\Var(U)=1/4-w\epsilon(1-\epsilon)\ge1/(4n)$. The construction of Appendix~\ref{app:common-posterior} and Lemma~\ref{lem:common-posterior-ws} give a finite structure $\theta_\epsilon\in\Theta_{\WS}$ with $M_i=U$ and $M=Y$. All experts report $[0,1]$ if and only if $U\in\{\epsilon,1-\epsilon\}$. This event happens with probability $w$, and conditional on this event, \(Y\) is $\mathrm{Bernoulli}(1/2)$. Thus, writing $h=f([0,1],\ldots,[0,1])$, every rule satisfies
\[
 R(f,\theta_\epsilon)
 \ge w\left\{\frac{1}{2}h^2+\frac{1}{2}(h-1)^2\right\}\ge\frac w4.
\]
As $\epsilon\downarrow\beta$, this lower bound tends to
\[
 \min\left\{\frac14,\frac{n-1}{4n(1-\alpha^2)}\right\}.
\]
For any $\varepsilon>0$, choosing $\epsilon$ sufficiently close to $\beta$, independently of $f$, gives the stated lower bound for every rule. If $\alpha>1/\sqrt n$, some $\epsilon>\beta$ permits $w=1$, so one structure gives regret at least $1/4$.
\end{proof}

\begin{proof}[Proof of Theorem~\ref{thm:interval-lower}]
If $0<\alpha<\min\{1/2,1/\sqrt n\}$, Lemma~\ref{lem:shrinkage-lower} supplies the required structure for the first two branches of Equation~\eqref{eq:interval-value}. If $1/2\le\alpha<1/\sqrt n$, Lemma~\ref{lem:central-lower} gives the third branch. For $1/\sqrt n\le\alpha<1$, the same lemma gives the last branch, including the approximating sequence at equality and the single structure above the threshold. These cases cover every $0<\alpha<1$, and each chosen structure works simultaneously for all aggregation rules.
\end{proof}

\subsection{Proof of Theorem~\ref{thm:interval-optimality}}\label{app:interval-proof}

\begin{proof}
Proposition~\ref{prop:interval-upper} bounds the worst-case regret of $f^*_{n,\alpha}$ by the expression in Equation~\eqref{eq:interval-value}. Theorem~\ref{thm:interval-lower} bounds every rule's worst-case regret below by the same expression minus any $\varepsilon>0$, using binary structures within $\Theta_{\WS}$. Letting $\varepsilon\downarrow0$ and taking the infimum over rules proves Equation~\eqref{eq:interval-value} and optimality of the rule in Equation~\eqref{eq:optimal-rule}.

The first three branches of Equation~\eqref{eq:interval-value} are strictly below $1/4$, and adjacent expressions agree at their boundaries. At $\alpha=1/\sqrt n$, Proposition~\ref{prop:interval-upper} and Equation~\eqref{eq:weight-objective-pieces} give the upper bound $1/4$ for every $\eta\in[0,\min\{2/\sqrt n,1\}]$; the matching lower bound makes each of these weights optimal.
\end{proof}

%% file: appendices/simple_average.tex
\section{Simple Average Comparisons}\label{app:simple-average}

This appendix establishes the simple-average benchmarks in Figure~\ref{fig:regret-comparison}. All worst cases range over the bounded-outcome class $\Theta_{\WS}$, and regret is measured against the omniscient aggregator predicting the full-signal posterior mean $M=\E[Y\mid S]$.

\subsection{Simple Average of Posterior Means}\label{app:mean-sa}

\begin{proposition}[Simple average for posterior means]\label{prop:mean-sa}
When each expert reports $Z_i=M_i=\E[Y\mid S_i]$, the simple average $f_{\mathrm{mean}}^{\mathrm{SA}}(z)=n^{-1}\sum_i z_i$ is robust optimal, with
\[
 \begin{gathered}
 \inf_f\sup_{\theta\in\Theta_{\WS}}R(f,\theta)=\sup_{\theta\in\Theta_{\WS}}R(f_{\mathrm{mean}}^{\mathrm{SA}},\theta)
 =\frac{n-1}{4n}.
 \end{gathered}
\]
Here $f$ ranges over all measurable rules for aggregating the reported posterior means.
\end{proposition}

\begin{proof}
For each expert's posterior $W_i$, set $B_{i,t}=\mathbf1\{W_i>t\}$ and $h_t=\E B_{i,t}$. Then $\int_0^1h_t\,dt=\E W_i=m_i$, and the binary certificate in Lemma~\ref{lem:binary-to-bounded} holds with $\gamma=1$ and $\kappa=0$. Its regret bound gives
\[
 R(f_{\mathrm{mean}}^{\mathrm{SA}},\theta)\le\frac{n-1}{4n}.
\]

For the matching lower bound, apply the common-posterior information structure in Appendix~\ref{app:common-posterior} with common posterior mean $U$ distributed as follows:
\[
 \PP(U=0)=\PP(U=1)=\frac1{2n},
 \qquad \PP(U=1/2)=\frac{n-1}{n}.
\]
Then $\E U=1/2$ and $\Var(U)=1/(4n)$, so Lemma~\ref{lem:common-posterior-ws} gives a finite information structure $\theta\in\Theta_{\WS}$ with $M_i=U$ and $M=Y$. On the event $U=1/2$, the report vector is $(1/2,\ldots,1/2)$ and the conditional outcome variance is $1/4$. Thus every aggregation rule incurs regret at least $(n-1)/(4n)$.
\end{proof}

\subsection{Simple Average of Interval Midpoints}\label{app:interval-sa}

\begin{proof}[Proof of Corollary~\ref{cor:interval-sa}]
Write $f_1(z)=\bar c(z)$ for simple averaging of interval midpoints, and set $\beta=(1-\alpha)/2$.

\emph{Upper bound.} Taking $\eta=1$ in Lemma~\ref{lem:binary-coefficient} gives
\[
 \Gamma_\alpha(1)=
 \begin{cases}
 2/(1+\alpha),&0<\alpha\le1/2,\\
 1/(1-\alpha^2),&1/2<\alpha<1.
 \end{cases}
\]
Whenever $\Gamma_\alpha(1)\le n$, the binary certificate and threshold representation in Section~\ref{sec:interval-extension} satisfy Lemma~\ref{lem:binary-to-bounded} with $\gamma=\Gamma_\alpha(1)$ and $\kappa=0$. Hence, for every $\theta\in\Theta_{\WS}$ with bounded outcome,
\[
 R(f_1,\theta)\le\frac{n-1}{4n}\Gamma_\alpha(1).
\]
For $\alpha\le1/2$, the required condition holds because $\Gamma_\alpha(1)\le2\le n$. For $\alpha>1/2$, it also holds when $\alpha\le1/\sqrt n$.

A second certificate covers the remaining reporting levels. For a Bernoulli posterior with mean $m_i$ and the midpoint $c(z_i)$ of the reported interval, the three report categories in Equation~\eqref{eq:binary-report-regions} give
\[
\E_{Y\sim\mathrm{Bernoulli}(m_i)}[(Y-c(z_i))^2]=
 \begin{cases}
 m_i,& 0\le m_i\le\beta,\\
 1/4,& \beta<m_i\le1-\beta,\\
 1-m_i,& 1-\beta<m_i\le 1.
 \end{cases}
\]
Thus the binary squared loss is at most $\kappa=\max\{\beta,1/4\}$. Equivalently, its squared prediction bias is at most $-m_i(1-m_i)+\kappa$, which is the certificate in Lemma~\ref{lem:binary-to-bounded} with $\gamma=0$. The midpoint threshold representation in Equation~\eqref{eq:interval-threshold-representation} with $\eta=1$ lets us apply that lemma directly, yielding
\[
 R(f_1,\theta)\le\max\{\beta,1/4\}
\]
for every $\theta\in\Theta_{\WS}$ with bounded outcome.

For $0<\alpha\le1/2$, combining the two certificates gives
\[
 R(f_1,\theta)\le
 \min\left\{\frac{1-\alpha}{2},
             \frac{n-1}{2n(1+\alpha)}\right\}.
\]
For $1/2<\alpha<1$, the second certificate gives $1/4$. The first certificate improves this bound exactly when $\alpha<1/\sqrt n$, where it is valid, so
\[
 R(f_1,\theta)\le
 \min\left\{\frac14,\frac{n-1}{4n(1-\alpha^2)}\right\}.
\]

\emph{Matching lower bounds.} Reuse the distribution of $U$ in Equation~\eqref{eq:four-posterior-law} from the proof of Lemma~\ref{lem:central-lower} with
\[
 0<\epsilon<1/2,\qquad
 w=\min\left\{1,\frac{n-1}{4n\epsilon(1-\epsilon)}\right\}.
\]
Then $\E U=1/2$ and $\Var(U)=1/4-w\epsilon(1-\epsilon)\ge1/(4n)$. The construction of Appendix~\ref{app:common-posterior} and Lemma~\ref{lem:common-posterior-ws} therefore give a finite structure $\theta_\epsilon\in\Theta_{\WS}$ with $M_i=U$ and $M=Y$.

If $0<\alpha\le1/2$, choose $\epsilon<\beta$. All experts report $[0,0]$ when $U\in\{0,\epsilon\}$, and report $[1,1]$ when $U\in\{1-\epsilon,1\}$. Thus, as $\epsilon\uparrow\beta$,
\[
 \begin{aligned}
 R(f_1,\theta_\epsilon)&=w\epsilon\longrightarrow
 \min\left\{\frac{1-\alpha}{2},
             \frac{n-1}{2n(1+\alpha)}\right\}.
 \end{aligned}
\]
If $1/2<\alpha<1$, choose $\epsilon>\beta$. All experts report $[0,1]$ when $U\in\{\epsilon,1-\epsilon\}$, and report $[U,U]$ when $U\in\{0,1\}$. Thus, as $\epsilon\downarrow\beta$,
\[
 \begin{aligned}
 R(f_1,\theta_\epsilon)&=\frac w4\longrightarrow
 \min\left\{\frac14,\frac{n-1}{4n(1-\alpha^2)}\right\}.
 \end{aligned}
\]
Together with the upper bounds, the two limits prove Equation~\eqref{eq:midpoint-regret}.

\emph{Range over reporting levels.} Both expressions inside the minimum in the first branch of Equation~\eqref{eq:midpoint-regret} strictly decrease on $(0,1/2]$. The second branch is nondecreasing on $[1/2,1)$, increasing until it reaches $1/4$. The branches agree at $\alpha=1/2$, with value $\min\{1/4,(n-1)/(3n)\}$. This proves the infimum in Equation~\eqref{eq:midpoint-envelope}.

For the supremum, the first branch has limit $(n-1)/(2n)$ as $\alpha\downarrow0$ and lies strictly below this limit, while the second is at most $1/4\le(n-1)/(2n)$. This proves the supremum in Equation~\eqref{eq:midpoint-envelope}.
\end{proof}

\subsection{Simple Average of Medians}\label{app:median-sa}

\begin{proof}[Proof of Corollary~\ref{cor:median-sa}]
Taking $\eta=1$ in inequality~\eqref{eq:median-upper} gives the upper bound for simple averaging:
\[
 R(f_{\mathrm{med},1},\theta)\le\frac{n-1}{2n}
 \qquad\text{for every }\theta\in\Theta_{\WS}.
\]
For the matching lower bound, use the finite structures $\theta_\delta$ from the proof of Proposition~\ref{prop:median-lower}. Under these information structures, the only median-report vectors are $0^n$ and $1^n$, each with probability $1/2$. Simple average sets $h_0=0$ and $h_1=1$ in each case. Hence
\[
 R(f_{\mathrm{med},1},\theta_\delta)
 =\rho b
 \longrightarrow\frac{n-1}{2n}
 \qquad\text{as }\delta\downarrow0.
\]
The unique medians make this matching lower bound valid for every selection method.
\end{proof}